\documentstyle[preprint,prb,aps]{revtex}
\begin{document}
\draft
\tighten

\title{Spontaneous Inter-layer Coherence in Double-Layer Quantum-Hall Systems
I: Charged Vortices and  Kosterlitz-Thouless Phase Transitions}

\author{K.~Moon, H.~Mori\cite{byline}, Kun~Yang, S.M.~Girvin, A.H.~MacDonald}
\address{Department of Physics, Indiana University, Bloomington, IN~~47405\\}

\author{L.~Zheng}
\address{Department of Physics and Astronomy,
University of Kentucky, Lexington,
KY~~40506\\}

\author{D.~Yoshioka}
\address{Institute of Physics, College of Arts and Sciences, University of
Tokyo, Komaba, Meguroku Tokyo 153, Japan\\}

\author{Shou-Cheng Zhang}
\address{Department of Physics, Stanford University, Palo Alto, CA~~94305}

\date{\today}

\maketitle

\begin{abstract}

At strong magnetic fields double-layer two-dimensional-electron-gas
systems can form an unusual broken symmetry state with spontaneous
inter-layer phase coherence.   In this paper we explore the rich variety of
quantum and finite-temperature phase transitions associated with
this broken symmetry.  We describe the system using a pseudospin
language in which the layer degree-of-freedom is mapped to a
fictional spin $1/2$ degree-of-freedom.  With this mapping the
spontaneous symmetry breaking is equivalent to that of a
spin $1/2$ easy-plane ferromagnet.  In this language
spin-textures can carry a charge.  In particular, vortices carry
$\pm e/2$ electrical charge and vortex-antivortex pairs can be
neutral or carry charge $\pm e$.  We derive an effective low-energy
action and use it to discuss the charged and collective neutral
excitations of the system.  We have obtained the parameters of the
Landau-Ginzburg functional from first-principles estimates and from
finite-size exact diagonalization
studies. We use these results to estimate the dependence of the
critical temperature for the Kosterlitz-Thouless phase
transition on layer separation.

\end{abstract}


\newpage
\pagenumbering{roman}
\tableofcontents
\pagenumbering{arabic}

\newpage
\section{INTRODUCTION}
\label{sec:intro}

Technological progress has made it possible to
produce double-layer two dimensional electron gas systems
of extremely high mobility.
As illustrated schematically in Fig.(\ref{fig1}), these systems consist
of a pair of 2D electron gases separated by a distance $d$
so small ($d\sim 100$\AA ) as to be comparable
to the typical spacing between electrons in the same layer.
In a large magnetic field, strong correlations between the layers
have long been expected to lead to novel fractional quantum Hall
effects.  Correlations are especially important in the strong
magnetic field regime because all electrons can be
accommodated within the lowest Landau level and execute cyclotron
orbits with a common kinetic energy.
The fractional quantum Hall effect occurs when the
system has a gap for making charged excitations, {\it i.e.}
when the system is incompressible, and theory has
predicted\cite{halperinz1,gsnumamd,amdreview} that at some Landau level
filling factors, gaps occur in double-layer systems
only if interlayer interactions are sufficiently strong.
These theoretical predictions have recently been confirmed\cite{expamd}.
More recently work from several different points of
view\cite{wenandzee,ezawa,ahmz1,gapless,harfok} has suggested that
inter-layer correlations can also lead to unusual broken symmetry
states with spontaneous phase coherence between layers which
are isolated except for inter-layer Coulomb interactions.
We have recently argued\cite{usPRL} that it is spontaneous interlayer
phase coherence which is responsible for the recently
discovered\cite{murphyPRL} extreme sensitivity of the
fractional quantum Hall effect at total Landau level filling
factor $\nu = 1$ to small tilts of the magnetic field away from
the normal to the layers. ($\nu \equiv N / N_{\phi}$ where $N$ is the
number of electrons and $ N_{\phi}$ is the number of single-particle
levels per Landau level.)

We have previously\cite{usPRL} presented a phenomenological theory of
the rich zero-temperature phase diagram associated with
spontaneous interlayer coherence.  In the present paper
we provide a detailed microscopic derivation of the effective action
used in our phenomenological theory.  We also
discuss the low-energy neutral and charged
excitations of the system in some detail.  Much of our discussion of the
properties of double-layer systems with spontaneous interlayer
coherence will be couched in language based on a simple
mapping\cite{ahmz1} of the layer
degree-of-freedom in a double-layer system to an artificial
`pseudospin' degree-of-freedom.  In this language the
spontaneous-interlayer-coherence broken symmetry appears as
easy-plane ferromagnetism.  The mapping is convenient because
the Hamiltonian of the system may be simply expressed in terms
of pseudospin operators and because some aspects of the physics
are familiar when expressed in this way.  This mapping is
discussed in detail in Section \ref{sec:analogy}.
In Section \ref{sec:charge} we present a microscopic derivation
of the connection between spin-textures and Coulomb charges in our model.
This relationship was discussed previously by
Sondhi {\it et al.}\cite{sondhi} in the context of a Chern-Simons effective
field theory
for the case of a single-layer
system at $\nu=1$ with weak Zeeman coupling (to the real spin).
In Section \ref{sec:action} we derive an effective action which
describes the low-energy physics of the system whenever the
system has spontaneous interlayer coherence {\it and} is
incompressible.  We use the spin-charge connection
and our effective action in Section \ref{sec:vortices} to
discuss the low-lying excitations of the system which are
formed from vortices in the pseudospin configuration.  We show that
vortices carry charge $\pm e/2$ and that the vortices appear
in four flavors corresponding to the independently available
sign choices for vorticity and Coulomb charge.  Neutral excitations
of finite energy can be formed from vortex pairs with both
opposite vorticity and opposite charge.  Collective spin-wave-like
excitations of the system also occur and dominate response-functions
at long-wavelengths.  These collective modes are discussed
in Section \ref{sec:colmodes}.   We believe that a double-layer
system with spontaneous interlayer coherence will have a
finite-temperature Kosterlitz-Thouless phase
transition\cite{wenandzee,ezawa} which we
discuss in Section \ref{sec:kostthou}.  In the low-temperature
phase a kind of superconducting behavior will occur in which the linear
resistivity vanishes when opposite currents are carried in the two
layers.  Fully microscopic calculations for the double-layer
systems, using exact diagonalization
studies of finite-size systems and many-body perturbation theory,
are reported in
Section \ref{sec:exactdiag}.   These calculations allow us to
estimate the parameters of the low-energy effective action
and hence to provide quantitative estimates of the
dependence of the Kosterlitz-Thouless temperature on the
separation between the layers.  Section \ref{sec:chern-simons} gives a
brief summary of the Chern-Simons effective field theory description of
double layer systems.
Finally in Section
\ref{sec:summary} we briefly summarize our findings.
A companion paper\cite{II} will discuss issues which arise when a
weak symmetry-breaking tunneling term is added to the
Hamiltonian of the double-layer system, particularly
those issues which arise from the recent experiments of Murphy
{\it et. al.}\cite{murphyPRL}.

\section{SPIN ANALOGY}
\label{sec:analogy}

We wish to show that the double-layer system at certain total
filling factors, particularly at $\nu=1$, can be viewed as an
{\it easy-plane quantum itinerant ferromagnet}.  In this section we will
give a qualitative introduction to the essential ideas of the physical
picture.  The mathematical details of the microscopic physics will be
presented in the subsequent sections.

We will use a `pseudospin' magnetic
language in which pseudospin `up' (`down') refers to an electron
in the `upper' (`lower') layer.\cite{datta} Using this language
and building upon recent progress in understanding the case of single-layer
systems at $\nu =1$ with real spin \cite{leekane,sondhi} we will
explore the consequences of the mixing of charge and pseudospin degrees of
freedom and discuss the rich variety
of phase transitions controlled by temperature, layer separation,
layer charge imbalance, and in the companion paper\cite{II} we explore
the effects of tunneling between layers, and magnetic field
tilt angles.
The present section will be devoted to development of a physical picture
of the rather counterintuitive concept of spontaneous phase coherence
between the layers, which in the magnetic analogy corresponds to
spontaneous pseudospin magnetization.  Technical details of the microscopic
calculations on which this picture is based will be presented in the
subsequent sections.

It is helpful to begin study of the pseudospin analogy by reminding ourselves
of the unusual properties of a single layer system at $\nu=1$ in the limit
of zero Zeeman splitting (for the `real' spins).\cite{sondhi,early}
In the presence of Coulomb repulsion between the particles, Hund's rule
would suggest that
the system could lower its interaction energy by maximizing its total spin
since states with maximum total spin are symmetric under spin
exchange and hence the spatial
wave function is necessarily fully antisymmetric.
In an ordinary ferromagnet the Hund's rule tendency to
maximize the total spin is partially counteracted by the increase in
kinetic energy (due to the Pauli principle) that accompanies spin
polarization.  In the lowest Landau level however, the kinetic energy has
been quenched by the magnetic field and the system will spontaneously
develop 100\% polarization.  An explicit microscopic wave function
believed\cite{faith} to exactly describe the ground state of
$N$ electrons at $\nu=1$ is\cite{singlet}
\begin{equation}
\Psi = \Psi_{V}
|\uparrow\uparrow\uparrow\uparrow\uparrow\uparrow
\uparrow\uparrow\uparrow\uparrow\uparrow\uparrow\ldots \uparrow\rangle,
\label{eq:smg1}
\end{equation}
where $\Psi_{V}$ is a Vandermonde determinant
wave function\cite{halperinz1} of the form
\begin{equation}
\Psi_{V} \equiv \prod_{i<j} \left(z_i - z_j\right)
\prod_k \exp (- |z_k|^2/ 4 \ell^2),
\label{eq:halperin111}
\end{equation}
where $\ell \equiv (\hbar c /e B)^{1/2}$ is the quantized cyclotron
orbit radius of the lowest Landau level.
The first term in Eq.(\ref{eq:smg1})
is simply the Laughlin spatial wave function for the filled
Landau level and the second term indicates that every spin is up.  This
state has total spin $S = N/2$ and $S^z\Psi = (N/2)\Psi$.  Because Coulomb
interactions do not directly affect the spins (magnetism is caused by
Coulomb forces plus the Pauli principle, not magnetic forces!),
$[H,S^\mu] = 0$ and $\Psi$
is simply one of a total of $2S+1$ degenerate states, all with $S=N/2$.
The other states are simply created using the total spin lowering operator
$S^- \equiv \sum_{j=1}^N s_j^-$ which is itself fully symmetric under spin
exchange.
Since the exact ground state (at $S^z=S=N/2$)
is a single Slater (Vandermonde) determinant
the two-particle distribution function for these states is readily computed
\begin{equation}
g(|{\bf r}-{\bf r'}|) = 1-e^{-|{\bf r} - {\bf r'}|^2/2} .
\label{smg2}
\end{equation}
Here and in the rest of the paper we set the magnetic length, $\ell$, to unity.
One clearly sees, in this expression, the `exchange hole' which surrounds
each particle and lowers its Coulomb energy by an amount
\begin{equation}
E_{\rm x} = \frac{1}{2\pi}\int d^2 r\, \frac{e^2}{r}\, [g(r) - 1]
 = e^2 \sqrt{\pi/2}  \sim 64{\rm K}~\sqrt{B [\text T]},
\label{smg3}
\end{equation}
where the dielectric function of the semiconductor environment is
implicit, $B$[T] indicates the magnetic field in Tesla,
 and the numerical estimate in Kelvins is for the case of GaAs.
For excited states, the spin wave function is not fully symmetric and
thus the spatial wave function is not fully antisymmetric.  There is then
a finite amplitude for particles of opposite spin to approach each other
closely.  The low-lying
excited states are `magnons' and just as for the
ferromagnetic Heisenberg model on a lattice, the exact single magnon
excitations for this (itinerant) magnet can be found.  They are labeled
by a conserved momentum ${\bf k}$ and have the form\cite{rasolt}
\begin{equation}
\Psi_{\bf k} = \overline{S^-_{\bf k}}\Psi,
\label{smg4}
\end{equation}
where
\begin{equation}
\overline{S^-_{\bf k}} \equiv \sum_{j=1}^N
\overline{{e^{i{\bf k\cdot r}_j}}} s^-_j,
\label{smg5}
\end{equation}
is the Fourier transform of the local spin lowering operator and the overbar
indicates projection onto the spatial wave functions of the lowest Landau
level.  The dispersion of these excitations is similar (at long
wavelengths) to those of
the Heisenberg model on a lattice and is given by\cite{rasolt,kallin}
\begin{equation}
\omega(k) =  \int {d^2 q \over (2 \pi)^2 } V(q) \exp (- |q|^2
/2 ) \left[1 - \cos( {\bf \hat z \cdot q \times k)} \right],
\label{smg6}
\end{equation}
where $V(q)$ is the Fourier transform of the electron-electron
interaction.
For large wave vectors one can show that this excitation crosses over
from being a collective spin wave mode to a single-particle type excitation
consisting of a  magnetic `exciton' which is a bound
state of a spin-flipped particle-hole pair.\cite{kallin}

In the discussion above, we have focussed on the ground state which has
definite total $S$ and have examined a basis in which $S^z$ is a good
quantum number.  It is convenient for later use however to recall that
since the Hamiltonian is invariant under spin rotations, we could have chosen
the spin quantization axis along any direction.  Suppose for example we
considered the ferromagnetic state $|\Psi_\varphi\rangle$ in which
every electron has the spinor
\begin{equation}
\left(\begin{array}{c} 1 \\ e^{i \phi}
\\ \end{array} \right).
\label{eq:spinorsmg}
\end{equation}
This state is an exact ground state of the Hamiltonian but
has indefinite $S^z$; that is, it is made up of a linear combination
of all the degenerate $S^z$ eigenstates.   The mean value of $S^z$ is
zero
\begin{equation}
\langle\Psi_\varphi | S^z | \Psi_\varphi\rangle = 0,
\label{smg7}
\end{equation}
but,
\begin{equation}
\langle \Psi_\varphi|{\bf S}|\Psi_\varphi\rangle =
\frac{N}{2}\left[\cos(\varphi) {\bf \hat x} + \sin(\varphi)
 {\bf\hat y}\right],
\end{equation}
showing that the spin is fully aligned but now lies in the xy plane.
Fluctuations in $S^z$, while non-zero, are relatively small (in the limit of
large $N$)
\begin{equation}
\left[\langle\Psi_{\varphi} | \left[S^z\right]^2 | \Psi_{\varphi}
\rangle\right]^{1/2}
= \frac{1}{2}N^{1/2}.
\label{smg8}
\end{equation}
This state can be represented as a coherent superposition of the
eigenstates of $S^z$ (obeying $S^z|m\rangle = m |m\rangle$).  For large
$N$, we have, to a good approximation
\begin{equation}
|\Psi_\varphi\rangle = \sum_{m = -N/2}^{N/2} e^{-\frac{2}{N}m^2}
e^{-im\varphi} |m\rangle.
\end{equation}
Such a state has a coherence which is analogous to that in the BCS state
of a superconductor (with $S^z$ playing the role of number operator).
There is a definite phase relationship between states with different values
of $S^z$.  Notice that this is not a direct consequence of the dynamics,
but merely a result of our choice of linear combination of the degenerate
basis
vectors.  That is, there are no terms in the Coulomb
Hamiltonian which flip spins
and yet in the ground state
there can be a definite phase relationship between amplitudes for
different numbers of flipped spins.

We turn now from the application of these ideas to single-layer systems
with `real' spin to the analogous ideas for double-layer systems described
by pseudospin.  We will ignore `real' spin, assuming it to be frozen out by
the Zeeman energy, although this is not necessarily a valid assumption
at low B fields.  The spinors
\begin{equation}
 \left(\begin{array}{c} 1 \\ 0 \\ \end{array} \right),
 \left(\begin{array}{c} 0 \\ 1 \\ \end{array} \right)
\end{equation}
describe states in which the electron is in the upper or lower layer
respectively.  Thus the layer number difference operator is simply the
$z$ component of the total pseudospin
\begin{equation}
N_\uparrow - N_\downarrow = 2 S^z.
\end{equation}
One's first reaction upon thinking about pseudospin is that it is a perfectly
sensible concept as long as it is an Ising-like variable; i.e., each
electron is either in the upper layer or in the lower, but not both.
However it is a fundamental feature of quantum mechanics that it is
perfectly sensible to talk about states in which there is a coherent
superposition of two amplitudes and the layer index is therefore uncertain.
For certain filling factors\cite{wenandzee,ezawa,ahmz1,gapless,harfok}
the ground state spontaneously develops interlayer
coherence and the electron pseudospin condenses into a state
\begin{equation}
\alpha_
\varphi = \left(\begin{array}{c} 1 \\ e^{+i\varphi}
\\ \end{array} \right)
\label{eq:spinorsmg3}
\end{equation}
magnetized in the xy plane.  Such a state has $\langle N_\uparrow -
N_\downarrow \rangle = 0 $ which reduces the charging energy of the double
layer system.  Such a state also has good exchange energy because, if two
electrons of the same pseudospin orientation (phase $\varphi$) approach
each other, the spatial part of the wave function must vanish.  It is this
exchange effect which gives rise to the finite pseudospin `stiffness'
which polarizes the pseudospins.  Of course, in the absence of tunneling,
$Q_- \equiv N_\uparrow - N_\downarrow$ is a good quantum number while
our variational wave function $\Psi_\varphi$ has $N^{1/2}$ fluctuations
in $Q_-$.  In analogy with coherent BCS states however, this is not
important to the physics (usually), it is simply mathematically
convenient not to project $\Psi_\varphi$ onto a state of definite
$Q_-$.  Of course it is perfectly possible to do so using
\begin{equation}
\Psi_{Q_-} = \int d\varphi e^{-iQ_-\, \varphi} \Psi_\varphi,
\end{equation}
without seriously modifying the good correlations built into the wave
function.

Even though the total number of electrons in a given
layer is a good quantum number, the dynamics will enforce a
definite phase relationship among states with different $S^z$ (layer charge
difference) due to a spontaneously broken U(1)
symmetry\cite{wenandzee,ezawa,harfok}
corresponding to rotations in pseudospin space about the ${\hat{\bf z}}$ axis.
This is quite analogous to what happens in superconductors.  For finite
layer separation, the charging energy will limit the fluctuations in
$Q_-$ and modified correlations will have to be built into the wave
function as discussed in Section \ref{sec:colmodes}.

The above discussion for $\nu = 1$ is, of course, somewhat over-simplified
for general filling factors.
Hund's rule, which suggests that the ground state should
have the maximum total spin quantum number consistent with
the Pauli exclusion principle, does not always apply to two-dimensional
electrons in the strong magnetic field limit.  In particular it is
known both from theoretical work\cite{halperinz1,rasolt,singlettheory}
and from experimental work\cite{singletexp} that at some filling factors
incompressible ground states can occur which are spin
singlets.\cite{singletcomment}
We focus our discussion here on the case $\nu=1$ where
the consequences of spontaneous-interlayer-coherence are
likely to be most easily observable.  At this filling factor there
is ample evidence that Hund's rule does give the correct answer for
the spin quantum number of the ground state.   We believe that much of the
physics we discuss will occur at any filling factor for which the
corresponding spin system has an incompressible ground state which
{\it is not} a spin-singlet.  In particular, the ground state for
$\nu =1/m$ for any odd integer $m$ is believed to have $S=N/2$
for any physically realistic interaction.  The ground state orbital
wavefunctions at these filling factors are well approximated by the
simple Jastrow wavefunctions discovered by Laughlin\cite{laughlin,Laughlin83}
and we will present some results for $m=3$ and $m=5$ based on these
trial wavefunctions.

\section{Spin-Charge Relation}
\label{sec:charge}
\subsection{Review of projection onto the lowest Landau level}

A convenient formulation of quantum mechanics within the subspace of the
lowest Landau level (LLL) was developed
by Girvin and Jach \cite{GJ}, and was exploited by Girvin, MacDonald and
Platzman in the magneto-roton theory of collective excitations of the
incompressible states responsible for the fractional quantum Hall
effect
\cite{GMP}.  Here we briefly discuss the part of this formalism that is
most relevant to the present paper.

We first consider the one-body case and choose the symmetric gauge.
The single-particle eigenfunctions of kinetic energy and angular momentum
in
the LLL are \cite{Laughlin83,GJ}
\begin{equation}
\phi_m(z)=\frac{1}{(2\pi 2^m m!)^{1/2}}\> z^m\>
 \exp{\biggl( -\frac{\vert z\vert^2}{4}\biggr)} ,
\label{eq3.10}
\end{equation}
where $m$ is a non-negative integer, and $z = (x + iy)/\ell$.  From
(\ref{eq3.10}) it is clear that any
wave function in the LLL can be written in the form
\begin{equation}
\psi (z)=f(z)\> e^{-\frac{\vert z\vert^2}{4}}
\label{eq3.20}
\end{equation}
where $f(z)$ is an analytic function of $z$,  so the subspace in the LLL is
isomorphic to the Hilbert space of analytic
functions \cite{Bargmann,GJ,stonebook}.
Following Bargmann \cite{Bargmann,GJ}, we define the inner product of two
analytic functions as
\begin{equation}
(f, g)=\int d\mu (z)\> f^\ast (z)\> g(z),
\label{eq3.30}
\end{equation}
where
\begin{equation}
d\mu (z)\equiv (2\pi )^{-1}\> dxdy\> e^{-\frac{\vert z\vert^2}{2}} .
\label{eq3.40}
\end{equation}

Now we can define bosonic ladder operators that connect $\phi_m$ to
$\phi_{m\pm
1}$ (and which act on the polynomial part of $\phi_m$ only):
\begin{mathletters}
\label{eq3.50}
\begin{equation}
a^\dagger = {z\over \sqrt{2}} ,\label{eq3.50a}
\end{equation}
\begin{equation}
a = \sqrt{2}\> \frac{\partial}{\partial z} ,\label{eq3.50b}
\end{equation}
\end{mathletters}
so that
\begin{mathletters}
\label{eq3.60}
\begin{equation}
a^\dagger\> \varphi_m = \sqrt{m+1}\> \varphi_{m+1} ,\label{eq3.60a}
\end{equation}
\begin{equation}
a\> \varphi_m = \sqrt{m}\> \varphi_{m-1} ,\label{eq3.60b}
\end{equation}
\begin{equation}
(f, a^\dagger\; g) = (a\; f, g) , \label{eq3.60c}
\end{equation}
\begin{equation}
(f, a\; g) = (a^\dagger\; f, g) .\label{eq3.60d}
\end{equation}
\end{mathletters}

\noindent All operators that have non-zero matrix elements only within the
LLL can be
expressed in terms of $a$ and $a^\dagger$.  It is essential
to notice that the
adjoint of $a^\dagger$ is not $z^\ast/\sqrt{2}$ but $a\equiv
\sqrt{2}\partial/\partial z$,
because
$z^\ast$ connects states in the LLL to higher Landau levels. Actually $a$ is
the
projection of $z^\ast/\sqrt{2}$ onto the LLL as seen clearly in the
following expression:
\[(f, \frac{z^\ast}{\sqrt{2}}\;
g)=(\frac{z}{\sqrt{2}}\; f, g)=(a^\dagger\; f, g)=(f, a\; g).\]
So we find
\begin{equation}
\overline{z^\ast}=2\frac{\partial}{\partial z},
\label{eq3.70}
\end{equation}
where the overbar indicates projecton onto the LLL.
Since $\overline{z^\ast}$ and $z$ do not commute, when we need to project
an
operator which is a combination of $z^\ast$ and $z$, we must first
normal
order $z^\ast$'s to the left of $z$'s, and then replace $z^\ast$ by
$\overline{z^\ast}$.  With this rule in mind and (\ref{eq3.70}), we
can easily project onto the LLL any operator
that involves space coordinates only.

  For example, the one-body density operator in momentum space is
\[
\rho_{\bf q} = \frac{1}{\sqrt{A}}\> e^{-i{\bf q}\cdot {\bf r}} =
\frac{1}{\sqrt{A}}\> e^{-\frac{i}{2}(q^\ast z + qz^\ast)} =
\frac{1}{\sqrt{A}}\>
e^{-\frac{i}{2}qz^\ast}\; e^{-\frac{i}{2}qz} ,
\]
where $A$ is the area of the system, and $q=q_x + iq_y$. Hence
\begin{equation}
\overline{\rho_q} = \frac{1}{\sqrt{A}}\> e^{-iq\frac{\partial}{\partial z}}\;
e^{-\frac{i}{2}q^\ast z} = \frac{1}{\sqrt{A}}\> e^{-\frac{\vert
q\vert^2}{4}}\;
\tau_q ,
\label{eq3.80}
\end{equation}
where
\begin{equation}
\tau_q = e^{-iq\frac{\partial}{\partial z} - \frac{i}{2}q^\ast z}
\label{eq3.90}
\end{equation}
is a unitary operator satisfying the closed Lie algebra
\begin{mathletters}
\label{eq3.100}
\begin{equation}
\tau_q\tau_k = \tau_{q+k}\> e^{\frac{i}{2}q\wedge k} ,\label{eq3.100a}
\end{equation}
\begin{equation}
[\tau_q, \tau_k] = 2i\; \tau_{q+k}\> \sin{\frac{q\wedge k}{2}}
,\label{eq3.100b}
\end{equation}
\end{mathletters}

\noindent where $q\wedge k \equiv \ell^2({\bf q}\times {\bf k})\cdot
 {\hat{\bf z}}$.  We also have
$\tau_q\tau_k\> \tau_{-q}\tau_{-k} = e^{iq\wedge k}$.  This is a
familiar feature
of the group of
translations in a magnetic field, because $q\wedge k$ is exactly the phase
generated by the flux in the parallelogram generated by ${\bf q}\ell^2$ and
${\bf k}\ell^2$.  Hence the $\tau$'s form a representation of the magnetic
translation group [see Fig.(\ref{fig:magtrans})].
In fact $\tau_q$ translates the particle
a distance $\ell^2\hat{{\bf z}}\times{\bf q}$. This means that different wave
vector components of the charge density do not commute. It is from here
that non-trivial dynamics arises even though the kinetic energy is totally
quenched in the LLL subspace.

This formalism is readily generalized to the case of many particles with
spin, as we will show next.  In a system with area $A$ and $N$ particles
the
projected charge and spin density operators are
\begin{mathletters}
\label{eq3.110}
\begin{equation}
\overline{\rho
_q} = \frac{1}{\sqrt{A}}\> \sum_{i=1}^N \overline{e^{-i{\bf
q}\cdot
{\bf r}_i}} = \frac{1}{\sqrt{A}}\> \sum_{i=1}^N e^{-\frac{\vert
q\vert^2}{4}}\> \tau_q(i)
\label{eq3.110a}
\end{equation}
\begin{equation}
\overline{S_q^\mu} = \frac{1}{\sqrt{A}}\> \sum_{i=1}^N \overline{e^{-i{\bf
q}\cdot {\bf r}_i}}\> S_i^\mu = \frac{1}{\sqrt{A}}\> \sum_{i=1}^N
e^{-\frac{\vert q\vert^2}{4}}\> \tau_q(i)\> S_i^\mu ,
\label{eq3.110b}
\end{equation}
\end{mathletters}

\noindent where $\tau_q(i)$ is the magnetic translation operator for the $i$th
particle
and $S_i^\mu$ is the $\mu$th component of the spin operator for the $i$th
particle.  We
immediately
find that unlike the unprojected operators, the projected spin and charge
density operators do not commute:
\begin{equation}
[\bar{\rho}_k, \bar{S}_q^\mu] = \frac{2i}{\sqrt{A}}\> e^{\frac{\vert k +
q\vert^2 -
\vert k\vert^2 - \vert q\vert^2}{4}}\> \overline{S_{k + q}^\mu}\>
\sin{\biggl(\frac{k\wedge q}{2}\biggr)} \neq 0.
\label{eq3.120}
\end{equation}
This implies that within the LLL, the dynamics of spin and charge are
entangled, i.e., when you rotate spin, charge gets moved.  As a
consequence of
that, spin textures carry charge\cite{sondhi}, as we will soon see.

\subsection{Spin-charge relation within the lowest Landau level}
We have argued above that in the SU(2) invariant case (i.e.,
when $d/\ell = 0$), the ground state at $\nu=1/m$ has its
pseudospin fully polarized spontaneously. At these filling factors the
interaction energy is minimized in such a state.
For the same reason, we might expect that the low-lying excited states
will be spin textures in which the local spin alignment varies slowly
with position.  To be explicit, we define the following as a
spin texture state:
\begin{equation}
\vert\tilde{\psi}[{\bf m}({\bf r})]\rangle
= e^{-i\bar{O}}\> \vert\psi_0\rangle .
\label{eq3.130}
\end{equation}
Here $\vert\psi_0\rangle$ is the $S^z=N/2$ member of the
ground state spin-multiplet given in Eq.(\ref{eq:smg1})
and the operator $O$
is a non-uniform spin rotation which reorients the local spin direction
from $\hat{\bf z}$ to ${\bf m}({\bf r})$ (${\bf m}$ is a unit vector).
We limit ourselves to small tilts away from the ${\bf\hat z}$ direction
so that
\begin{equation}
O = \sum_{j=1}^N {\bf\Omega}({\bf r}_j) \cdot {\bf S}_j \equiv \sum_q
e^{\frac{\vert
q\vert^2}{4}}\> \Omega_q^\mu\> S_{-q}^\mu ,
\label{eq3.140}
\end{equation}
where
${\bf \Omega}({\bf r}) = {\bf\hat{z}}\times {\bf m}({\bf r})$ is the angle
over which a spin is rotated. [Note that $\Omega^z({\bf r}) \equiv 0$,
$\Omega^x({\bf r}) = - m^y({\bf r})$, and $\Omega^y({\bf r}) = m^x(
{\bf r})$].  We will later argue that our final result requires only
that $\Omega$ is slowly varying in space and not
that $\Omega$ is small.  However the assumption is convenient at the
present since it allows us to use a simple expression for $\Omega$
and also to expand $\bar O$ as a small quantity.  Projecting $O$ onto the LLL
ensures that $\vert\tilde{\psi}\rangle$ has no projection on higher
Landau levels \cite{footnote3.1} as required in the strong perpendicular
magnetic field limit.
The extra factor $e^{\frac{\vert q\vert^2}{4}}$
in Eq.(\ref{eq3.140}) implies a non-standard definition for
the Fourier components of $\Omega^{\mu} ({\bf r})$ which
is adopted as a convenience.

We can now calculate the excess charge density in a spin texture
state:
\begin{equation}
\delta \rho_k = \langle \psi_0 \vert e^{i\bar{O}}\>
\bar{\rho}_k\> e^{-i\bar{O}}
\vert
\psi_0 \rangle - \langle \psi_0 \vert \bar{\rho}_k \vert \psi_0 \rangle .
\label{eq3.150}
\end{equation}
Expanding in powers of $\bar{O}$ gives
\begin{equation}
\delta \rho_k = i\> \langle \psi_0 \vert [\bar{O}, \rho_k] \vert \psi_0
\rangle -
\frac{1}{2}\> \langle \psi_0 \vert [\bar{O}, [\bar{O}, \rho_k]] \vert \psi_0
\rangle + \dots .
\label{eq3.160}
\end{equation}
It is easy to check that the first term is zero.  Using (\ref{eq3.120}) we
obtain
\begin{equation}
[\bar{O}, \bar{\rho}_k] = \frac{2i}{\sqrt{A}}\> \sum_q e^{\frac{\vert k -
q\vert^2
- \vert k\vert^2 }{4}}\> \Omega_q^\mu\>
\overline{S_{k-q}^\mu}\> \sin{\frac{k\wedge q}{2}} ,
\label{eq3.170}
\end{equation}
substituting this into (\ref{eq3.160}), and keeping only the second-order term
we obtain
\widetext
\begin{equation}
\delta \rho_k = -\frac{i}{\sqrt{A}}\> \sum_{p,q}
e^{\frac{\vert k - q\vert^2 - \vert k\vert^2 +|p|^2}{4}}\>
\Omega_q^\mu\> \Omega_p^\nu\>
\sin{\frac{k\wedge q}{2}}\> \langle \psi_0 \vert [\overline{S_{-p}^\nu},
\overline{S_{k-q}^\mu}] \vert \psi_0 \rangle .
\label{eq3.180}
\end{equation}
\narrowtext
At this point we use our assumption that
$\vert \psi_0 \rangle$ is a state with uniform
spin density so that the  expectation value in Eq.(\ref{eq3.180}) is
non-zero only when $p=k-q$.  Using
\begin{equation}
[\overline{S_{q-k}^\nu}, \overline{S_{k-q}^\mu}] = -\frac{i}{A}\>
e^{-\frac{\vert k-q\vert^2}{2}}\> \epsilon_{\mu\nu\lambda}\> \sum_{j=1}^N
S_j^\lambda ,
\label{eq3.190}
\end{equation}
we obtain (using the fact that only the $z$ component of spin is non-zero)
\begin{equation}
\delta \rho_k = -\frac{N}{2A^{3/2}}\> \epsilon_{\mu\nu}\> e^{-\frac{\vert
k\vert^2}{4}}\> \sum_q \Omega_q^\mu\> \Omega_{k-q}^\nu\> \sin{\frac{k\wedge
q}{2}} .
\label{eq3.200}
\end{equation}
Since $\Omega$ is a slowly varying function, $\Omega_q^\mu$ is negligible
when
$q$ is large, we can make the approximation
\begin{equation}
e^{-\frac{\vert k\vert^2}{4}}\> \sin{\frac{k\wedge q}{2}} \simeq
\frac{k\wedge
q}{2} .
\label{eq3.210}
\end{equation}
Substituting into (\ref{eq3.200}) we obtain
\begin{eqnarray}
\delta \rho_k &=& -\frac{\nu}{8\pi}\> \frac{1}{\sqrt{A}}\> \epsilon_{\mu\nu}\>
\sum_q \Omega_q^\mu\> \Omega_{k-q}^\nu\> (k\wedge q)\nonumber\\
&=& -\frac{1}{8\pi}\> \frac{\nu}{\sqrt{A}}\>
\epsilon_{\mu\nu}\>
\sum_q (iq\> \Omega_q^\mu )\wedge (i(k-q)\> \Omega_{k-q}^\nu )\nonumber\\
&=& -\frac{1}{8\pi}\> \frac{\nu}{\sqrt{A}}\>
\epsilon_{\mu\nu}\>
\sum_q (\nabla\Omega^\mu )_q\wedge (\nabla\Omega^\nu )_{k-q}\nonumber\\
           &=& -\frac{\nu}{8\pi}\> \epsilon_{\mu\nu}\> \Biggl\{
(\nabla\Omega^\mu )\wedge (\nabla\Omega^\nu )\Biggr\}_k .
\label{eq3.220}
\end{eqnarray}
Here we have used the fact that $N/A = \nu / (2 \pi)$.
Fourier transforming back to real space, we obtain
\begin{equation}
\delta \rho({\bf r}) = -\frac{\nu}{8\pi}\> \epsilon_{\mu\nu}\>
\biggl(\nabla\Omega^\mu ({\bf r})\wedge \nabla\Omega^\nu ({\bf r})\biggr) .
\label{eq3.230}
\end{equation}
Expressing $\delta \rho$ in terms of ${\bf m}$ instead of ${\bf\Omega}$, we
finally obtain
\begin{equation}
\delta \rho({\bf r}) = -\frac{\nu}{8\pi}\> \epsilon_{\mu\nu}\>
{\bf m}({\bf r})\cdot
\left[\partial_\mu {\bf m}({\bf r})\times
\partial_\nu {\bf m}({\bf r})\right]
\label{eq:3.240}
\end{equation}
which is exactly $\nu$ times the
Pontryagian index density, or topological charge
density.\cite{sondhi,fradkin}
In Eq.(\ref{eq:3.240}) we have used the fact that $\Omega$ is small to replace
${\hat{\bf z}}$ by ${\bf m}$.
  The final result depends on $\nabla {\bf\Omega}$
instead of ${\bf\Omega}$ and it is clear\cite{P6} that the expansion
in (\ref{eq3.160}) is actually an expansion in terms of $\nabla {\bf\Omega}$,
rather than $\Omega$. Hence our final result is
valid as long as  ${\bf\Omega}$ is slowly varying so that
$\nabla {\bf\Omega}$ is small (compared to $\ell^{-1}$).
\cite{sondhi,fradkin}

The density in Eq.(\ref{eq:3.240}) can be viewed as the time-like component
of a conserved (divergenceless) topological `three-current'
\begin{equation}
j^\alpha = -\frac{\nu}{8\pi}\> \epsilon^{\alpha\beta\gamma}\>
\epsilon_{abc}\>
{m^a}({\bf r})
\partial_\beta {m^b}({\bf r})
\partial_\gamma {m^c}({\bf r}).
\label{eq:3.240a}
\end{equation}
Using the fact that $\bf m$ is a unit vector, it is straightforward to
verify that $\partial_\mu j^\mu = 0$.

Thus we have shown that for spin-states with
$S=N/2$ the physical charge density is $\nu$ times the
the topological charge density, in the long wavelength limit.
This remarkable result was first obtained by Sondhi {\em et} {\em al}.
within the context of a Chern-Simons effective field theory
description of spin textures.
\cite{sondhi} The present derivation gives a microscopic proof
of their result.
The total extra charge carried by the spin texture is exactly the
Pontryagian index:
\begin{equation}
\triangle N=-\frac{\nu}{8\pi}\>
\int{d^2{\bf r}}\, \epsilon_{\mu\nu}\> {\bf m}({\bf r})\cdot
\left[\partial_\mu {\bf m}({\bf r})\times \partial_\nu {\bf m}({\bf
r})\right],
\label{eq3.245}
\end{equation}
which is an integer multiple of $\nu$ because it is the number of times
a unit sphere is wrapped around by the order parameter, i.e., it is
the winding number of the spin texture.\cite{fradkin}
For $\nu =1/m $ and  $m = 3,5 $ elementary spin-textures
carry the same fractional charge
as the quasiparticles discovered by Laughlin\cite{laughlin}
for spinless electrons.  As we discuss below the fact that the
charges are the same follows from very general considerations.
Actually the spin texture states we have defined must contain
precisely the same number of particles as $|\psi_0 \rangle$ since
the spin-rotation operator does not change the total electron
number.  However the spin-density may contain a number of well-separated
textures with well-defined non-zero topological charge densities and
hence well localized charges; only the net charge in the spin-texture
states defined above will be zero.  The system clearly also has states
with locally non-zero net charge in the spin textures.

The fact that spin textures carry charge can also be understood from
the following very different point of view. In the Hartree-Fock picture
the electrons see a very strong exchange field which locally aligns
the spins ferromagnetically to produce the spin texture.
As an electron propagates through
this (slowly spatially varying) exchange field, its spin adiabatically
follows the local orientation of the exchange field.
A consequence of the tilting of the spin is that when an electron moves along
a closed path that surrounds the area $\Gamma$, the spin contributes
a Berry's phase to the path integral\cite{fradkin}:
\begin{equation}
\varphi = -\frac{1}{4}\> \int_{\Gamma}{d^2 {\bf r}}\,\epsilon_{\mu\nu}\>
{\bf m}({\bf r})\cdot
\left[\partial_\mu {\bf m}({\bf r})\times \partial_\nu
{\bf m}({\bf r})\right].
\label{eq3.250}
\end{equation}
This extra phase is exactly equivalent to having an Aharonov-Bohm phase
due to additional magnetic
flux inside $\Gamma$.  In our system, the Hall
conductance is not zero:
\begin{equation}
\sigma_{xy}={\nu e^2\over h},
\label{eq3.260}
\end{equation}
which means that additional flux $\Phi$ gives additional charge
$Q= e\nu \Phi/\Phi_0$.
\cite{kagome} This is the same mechanism that causes Laughlin
quasiparticles to carry quantized fractional charge when
$\sigma_{xy}$ is quantized.\cite{laughlin}
Combining (\ref{eq3.250}) and (\ref{eq3.260}) tells us that the additional
charge density is given by Eq.(\ref{eq:3.240}).\cite{kagome}

\section{Effective Action}
\label{sec:action}

\subsection{SU(2)-invariant interactions}
In this section we calculate the effective action of a smooth spin texture for
$SU(2)$-invariant (i.e.
$d/\ell =0$) electron-electron interactions.  The considerations
in this section apply to both a single-layer with zero Zeeman
energy and to a pseudospin-polarized double-layer system in which the
two layers are spatially coincident
so that the interactions between layers and within layers are the same.
In the double-layer case this limit can of course
never be achieved experimentally, but it is a convenient place to begin
the analysis since the ground state is exactly soluble in this limit.
We assume, for the sake of convenience, that the spins are almost
aligned in the ${\hat{\bf z}}$ direction,
and that they vary slowly in space, i.e., $\Omega_q$ is negligible when
$q\ell\geq 1$.  The interaction, after projection onto the LLL, is

\begin{equation}
\overline{V} = \frac{1}{2}\;\sum_q V_q\;
(\overline{\rho}_q\; \overline{\rho}_{-q}-N e^{-\frac {q^2}{2}}),
\end{equation}
where
$V_k=\int d^2 r\, V(r) e^{-i{\bf k\cdot r}}.$
The expectation value of the energy is
\begin{eqnarray}
\delta E &=&
\langle \psi_0\vert e^{i\overline{O}}\,[ \overline{V},
e^{-i\overline{O}}]
\vert\psi_0\rangle\nonumber \\
           &=& -i \langle\psi_0\vert
           [\overline{V},\, \overline{O}] \vert\psi_0\rangle -
           \frac{1}{2}\;
           \langle\psi_0\vert [\overline{O},\, [\overline{O},\,
           \overline{V}]]
           \vert\psi_0\rangle
                            + \cdots
\end{eqnarray}
Since $\Omega^z=0$, the leading term vanishes.  The second-order term gives
\begin{eqnarray}
\delta E &=& -\frac{1}{2}\; \langle\psi_0\vert
[\overline{O},\, [\overline{O},\, \overline{V}]]
\vert\psi_0\rangle\nonumber\\
&=& -\frac{1}{4}\; \sum_k V_k\; \langle\psi_0\vert [\overline{O},\,
[\overline{O},\, \overline{\rho}_k\, \overline{\rho}_{-k}]]
\vert\psi_0\rangle.
\label{eq:commutator}
\end{eqnarray}
\noindent
The commutators in Eq.(\ref{eq:commutator}) can be evaluated using
\begin{eqnarray}
[\overline{\rho}_k,\, \overline{O}] &=& \frac{1}{\sqrt{A}}\;
e^{-\frac{|k|^2}{4}}\;
\sum_j
    [\tau_k (j),\, \sum_q \Omega_q^\mu\; \overline{S}_{-q}^\mu ]\nonumber\\
                                    &=& \frac{2i}{A}\; \sum_{j,q}
e^{-\frac{k^2}{4}}\; \Omega_q^\mu\;
S_j^\mu
    \; \tau_{k-q} (j)\;  \sin{\frac{q\wedge k}{2}} .
\end{eqnarray}
Substituting this into the above expression, we obtain for small $q$
\begin{eqnarray}
\delta E &=& \frac{-N}{2A^2}\; \sum_k V_k\; \sum_q
        (\Omega_q^x \Omega_{-q}^x + \Omega_q^y \Omega_{-q}^y)\;
\frac{1}{4}
(q\wedge k)^2 h(k) \nonumber\\
&=& \frac {\rho_s^0}{2}\; \sum_q [(iq) \Omega_q^x\; (-iq) \Omega_{-q}^x +
(iq) \Omega_q^y\; (-iq) \Omega_{-q}^y]\nonumber\\
&\equiv & \frac {\rho_s^0}{2}\; \int d^{2}r\; [(\nabla \Omega^x)^2 + (\nabla
\Omega^y)^2] = \frac {\rho_s^0}{2}\; \int d^{2}r (\nabla {\bf m})^2\; .
\label{nonlinearsigmamodel}
\end{eqnarray}
The spin stiffness $\rho_s^0$, implicitly defined above, is related to
the pair correlation function of $|\psi_0\rangle$  by
\begin{equation}
\rho_s^0 = \frac{-\nu}{32 \pi^2} \int dk  k^3
V_k h(k)
\label{eq:stiffness}
\end{equation}
where the pair-correlation function $h(k) \equiv (\nu / 2 \pi)
\int d^2 {\bf r}
(g(r) - 1 ) \exp (- i {\bf k \cdot r}) $.  For $\nu = 1$,
$ | \psi_0 \rangle$ is known analytically and the pair correlation
function can be evaluated analytically; $ h(k) = - \exp ( - |k|^2/2)$.

In this calculation we have kept the lowest
order gradient terms only. The physical origin of the stiffness is the loss of
exchange and correlation energy when the spin orientation varies with position.
For the Coulomb interaction,
$\rho_s^0 = e^2/ (16\sqrt {2\pi} \epsilon \ell)
\sim (e^2 / \epsilon \ell) 2.49 \times 10^{-2}$  at $\nu=1$.  For
$\nu =1/3$ and $\nu = 1/5$ we have evaluated $\rho_s^0$
numerically using hypernetted-chain-approximation\cite{aers} correlation
functions and find $\rho_s^0 = (e^2/\epsilon\ell) \, 9.23 \times 10^{-4}$ and
$\rho_s = (e^2/\epsilon \ell) \, 2.34 \times 10^{-4}$ respectively.
For $\nu = 1$ this is exactly the coefficient of the gradient term which Sondhi
$et$ $al.$ \cite {sondhi} obtained by fitting the known\cite{kallin,rasolt}
long wavelength spin wave spectrum,
but here we obtain it from a $first$-$principles$ calculation.
The classical model defined by Eq.(\ref{nonlinearsigmamodel})
is called the $O(3)$ non-linear sigma model and has been studied in
great detail.\cite{rajaraman}  We note in passing that for the SU(2)
invariant case, the spin stiffness $\rho_s$ found here is exact.  Quantum
fluctuation corrections to the Hartree-Fock approximation
affect only higher gradient terms in the action.

Now let us see what happens at higher order in $\Omega$.  The third
term is
\begin{equation}
E^{(3)}={-i\over 2}\sum_{k}{V_{k}{1\over 3!}\langle\psi_0|[\overline
O,[\overline O,
[\overline O,\overline \rho_{k}\overline \rho_{-k}]]]|\psi_0\rangle}.
\end{equation}
We have an odd number of powers of $S^{x}$ or $S^{y}$
combined together, but the state is polarized in the ${\bf\hat{z}}$
 direction, so the
expectation value must be zero.  In general odd order terms are zero
by symmetry.  The next non-zero term appears at fourth order:
\begin{equation}
E^{(4)}={1\over 2}\sum_{k}{V_{k}{1\over 4!}\langle\psi_0|[\overline
O,[\overline O,
[\overline O,[\overline O,\overline \rho_{k}\overline
\rho_{-k}]]]]|\psi_0\rangle}.
\end{equation}
The nested commutators can be expanded as
\begin{eqnarray}
[\overline O,[\overline O,[\overline O,[\overline O,
\overline \rho_{k}\overline \rho_{-k}]]]] &=& \overline \rho_{k}[\overline
O,[\overline O,[\overline
O,[\overline O,\overline \rho_{-k}]]]]+4[\overline O,\overline
\rho_{k}][\overline O,[\overline O,[\overline
O,\overline \rho_{-k}]]]\nonumber\\
&& +6[\overline O,[\overline O,\overline \rho_{k}]][\overline O,[\overline
O, \rho_{-k}]]
+4[\overline O,[\overline O,[\overline O,\overline \rho_{k}]]][\overline O,
\overline \rho_{-k}]\nonumber\\
&& +[\overline O,[\overline O,[\overline
O,[\overline O,
\overline \rho_{k}]]]]\overline \rho_{-k}.
\label{eq:fourthorder}
\end{eqnarray}
We remark that for $\nu = 1$ the first and last terms (with $k\ne 0$)
give no contribution to $E^{(4)}$,
since the ground state is annihilated by $\bar \rho_k$.  (There
are no nonuniform density, $S_z=N/2$ states since we have a full Landau
level.)
The third order term contains the Hartree-interaction between
the charges of the spin-textures as can be recognized after
the following manipulations:
\begin{eqnarray}
{1\over 8}\sum_{k}{V_{k}\langle\psi_0|[\overline O,[\overline O,\overline \rho
_{k}]][\overline O,
[\overline O,\overline \rho_{-k}]]|\psi_0\rangle} &\approx& {1\over 8}
\sum_{k}{V_k}\langle
\psi_0|[\overline O,[\overline O,\overline
\rho_{k}]]|\psi_0\rangle\langle\psi_0|[\overline O,
[\overline O,\overline \rho_{-k}]]|\psi_0\rangle\nonumber\\
&=&{1\over 2}\sum_{k}{V_{k}\langle\delta\rho_{k}\rangle\langle\delta\rho_
{-k}\rangle}\nonumber\\&=&{1\over 2}\int\, \int{d^{2
}r d^{2}r'\, V({\bf r - r}\,')
\delta\rho({\bf r})\delta\rho({\bf r}\,')},
\end{eqnarray}
where $\delta\rho={\nu\over 8\pi}\epsilon_{ij}(\partial_{i}
{\bf m}\times\partial
_{j}{\bf m})\cdot{\bf m}$
is the topological charge density. In the first step above we
have used the fact that
$|\psi_0\rangle$ is an approximate eigenstate of
$[\overline O,[\overline O,\overline \rho_{-k}]]$, since $\Omega^z$ is zero
and even combinations of the spin operators $S^x$ and $S^y$
(i.e. $S^x\,S^x$, $S^x\,S^y$, $etc$) commute with $S^z$.  (For $\nu=1$
this maneuver is exact.)  Thus the third term in
Eq.(\ref{eq:fourthorder}) contains the direct Coulomb energy term.  If we
had a short range interaction, other terms would give contributions of the
same importance, however because of the long-range of the
Coulomb interaction, i.e.,
because $V_{k}\sim 1/k$ as $k\rightarrow 0$, one can show
that this term is the dominant fourth order term in the
spin-texture energy.   We take this to be the next leading term in the
spin-texture energy functional.

So far we have calculated the most relevant terms of the static energy of a
spin texture. The dynamics can be obtained by studying the equation of motion.
The quantum equation of motion is
\begin{eqnarray}
{dm^{\mu}_{q}\over dt}&=&{4 \pi \over \nu} e^{{|q|^2\over 4}}\langle
{d S^{\mu}_{q} \over dt} \rangle
=-{ 4 \pi i \over \hbar \nu} \langle\tilde{\psi}|[e^{{|q|^2\over
4}}S^{\mu}_{q},\overline V]
|\tilde{\psi}\rangle
\simeq -{ 4 \pi \over \hbar \nu } \langle\psi_0|[\overline
O,[e^{{|q|^2\over 4}}S_{q}^{\mu},\overline V]
]
|\psi_0\rangle\nonumber\\
&=&-{2 \pi \over \hbar \nu }{\delta\over
\delta\Omega_{-q}^{\mu}}\langle\psi_0|[\overline O,
[\overline O,\overline V]]|\psi_0\rangle={ 4 \pi \over \hbar \nu}
{\delta\over \delta\Omega_{-q}^{\mu}}E[ {\bf m}],
\label{eq:eqmotion}
\end{eqnarray}
where $E[{\bf m}]$ is the energy functional of the spin texture,
\begin{eqnarray}
E[{\bf m}]=\frac {\rho_s^0}{2}\; \int d^{2}r (\nabla{\bf m})^2\;
+\; {1\over 2}\int\, \int{d^{2
}r d^{2}r'\, V({\bf r}-{\bf r}\,')
\delta\rho({\bf r})\delta\rho({\bf r}\,')}.
\end{eqnarray}
If we include only the leading gradient term in the energy functional, an
approximation which is always valid for sufficiently slowly varying
spin-textures, we obtain
\begin{equation}
{d {\bf m_q} \over  d t } = {4 \pi \rho_s^0 q^2\over \hbar \nu}
 {\hat{\bf z}} \times  {\bf  m_q}
\label{eq:precession}
\end{equation}
The equation of motion has spin-wave solutions in which the
magnetization precesses around the ${\hat{\bf z}}$ direction with wavevector
$q$ and frequency $\hbar \omega = (4 \pi \rho_s^0 q^2 /  \nu)$.
This is precisely the energy of the long-wavelength spin-waves
of the system.\cite{kallin,rasolt}.
This equation of motion immediately leads to the following
effective Lagrangian:
\begin{equation}
L= { \nu \over 4 \pi}
\int d^{2}r {\bf A}[{\bf m}({\bf r})]\cdot\partial_{t}{\bf m}({\bf r})
-E[{\bf m}],
\label{monopole_action}
\end{equation}
where ${\bf A}$ is the vector potential of a
unit magnetic monopole\cite{haldane83,fradkin} in the spin space; i.e.,
$\nabla_{\bf m}\times{\bf A}={\bf m}$.
[For spins oriented close to the ${\bf\hat z}$ direction as in the
above discussions ${\bf A} \approx (1/2) (- m_y, m_x, 0) $].
The first term simply contributes to the action a geometric phase
proportional to the solid angle traced out
by the spin vector during its motion.  This is exactly the Berry's
phase for the spin and appears at the adiabatic level
as expected\cite{stone}.

We have now derived all the terms in the
effective Lagrangian of Sondhi $et$ $al$. \cite{sondhi} from a
first-principles calculation.

\subsection{Effective action for symmetry-breaking interaction }

In this section we derive an effective action suitable for
double-layer systems using the pseudospin analogy discussed in
Section \ref{sec:analogy}.  We assume\cite{noscatt}
that terms in the Hamiltonian
where electrons are scattered from layer to layer by
interactions can be neglected and that the two wells are identical.
We define
\begin{equation}
V^0 \equiv \frac{1}{2}\; (V^A_k+V^E_k)
\end{equation}
\begin{equation}
V^z_k \equiv \frac{1}{2}\; (V^A_k-V^E_k)
\end{equation}
where $V^A_k$ is the Fourier transform with respect to the
planar coordinate of the
(intr{\it a}layer) interaction potential between a pair of
electrons in the same layer and $V^E_k$ is Fourier
transform of the (int{\it e}rlayer) interaction potential between a pair of
electrons in opposite layers.   If we neglect the finite
thickness\cite{thickness} of the layers, $V^A_k=2 \pi e^2 / k$
and $V^E_k = \exp (-kd) V^A_k$.
The interaction Hamiltonian can then be separated into
a pseudospin-independent part with interaction $V^0$ and
a pseudospin-dependent part.  The pseudospin dependent term in
the Hamiltonian is
\begin{equation}
\overline V_{\rm sb}
=2\sum_{k}{V_{k}^{z}\overline S_{k}^{z}\overline S_{-k}^{z}}.
\end{equation}
Since $V^A_k >V^E_k$, this term produces an
easy-plane rather than an Ising anisotropy.  The pseudospin symmetry of
the Hamiltonian is reduced from $SU(2)$ to $U(1)$ by this term.
In addition, this term changes the quantum fluctuations in the
system since
it does not commute with the order parameter
\begin{equation}
[\overline V_{\rm sb}, S^\mu] \ne 0,
\end{equation}
where $\mu = x,y$.

The pseudospin texture energy due to the pseudospin-independent term in the
Hamiltonian can be calculated as discussed in the previous section.
In this section we calculate the contribution of the pseudospin-dependent
term in the Hamiltonian to the spin-texture energy.

In calculating the contribution of the pseudospin-dependent
term to the spin-texture energy we approximate the ground state
by the $S=N/2$ total pseudospin eigenstate which {\it is} the ground
state in the limit of zero layer separation.
We argue that the
form of the energy-functional we derive must remain valid
even when quantum fluctuations due to the pseudospin-dependent
terms in the Hamiltonian are present.  However, the coefficients
which appear in the energy-functional will be altered by
quantum fluctuations and the explicit expressions we derive
below are accurate only when the pseudospin-dependent interactions
are weak, {\it i.e.}, only when the layers are close together.
In Section \ref{sec:exactdiag} we will discuss estimates
obtained for the quantum fluctuation corrections to
these  coefficients from finite-size
exact diagonalization and many-body perturbation theory calculations.

It is convenient here to take the ground state
$|\psi_0\rangle$ to be spin polarized along the ${\bf\hat{x}}$ direction.
We first calculate the energy change associated with small
oscillations of the spin-texture away from
the ${\bf\hat x}$ direction.  The leading term vanishes just as in the
SU(2) invariant case.
To understand qualitatively the
physics contained in the pseudospin-dependent term in the
Hamiltonian we focus first on the second-order term
\begin{equation}
\delta E_{\rm sb}=
-\sum_{k}V_{k}^{z}\langle\psi_0|2[\overline O,\overline
S_{k}^{z}][\overline O,\overline S_{-k}
^{z}]+\overline S_{-k}^{z}[\overline O,[\overline O, \overline S_{k}^{z}]]+
[\overline O,[\overline O, \overline
S_{-k}^{z}]]\overline S_{k}^{z}|\psi_0\rangle.
\label{eq:sbeng}
\end{equation}
\noindent
The first term on the right-hand
side of Eq.(\ref{eq:sbeng}) yields exactly (for $\nu = 1$)
\begin{eqnarray}
-2\sum_{k}V_{k}^{z}\langle\psi_0|[\overline O,\overline
S_{k}^{z}]|\psi_0\rangle
\langle\psi_0|[\overline O,\overline S_{-k}^{z}]|\psi_0\rangle
&=&{\nu^{2}\over 2A^{2}}\sum_{k}V^{z}_{k}e^{-\frac{k^2}{2}}
\Omega^{y}_{k}\Omega^{y}_{-k}\nonumber\\
&=&{\nu^{2}\over 2A^{2}}\sum_{k}V^{z}_{k}
e^{-\frac {k^2}{2}}m^{z}_{k}m^{z}_{-k}\nonumber\\
&=&{N^{2}\over 2A^{2}}\int{d^{2}r d^{2}r' \tilde{V}^{z}({\bf r}
-{\bf r}\,')m^{z}({\bf r})m^{z}({\bf r}\,')},
\end{eqnarray}
where $\tilde{V}(r)$ is the Fourier transform
of $V_{k}^{z}e^{-\frac{k^2}{2}}$. This
is exactly the Hartree-like charging energy. In the limit of a smooth
spin texture, using the gradient expansion, this term becomes a local mass
term, which is the capacitive charging energy. That is,
\begin{equation}
\beta_{H}\int{d^{2}r
(m^{z})^{2}}\,\,
\end{equation}
where
\begin{equation}
\beta_{H}= \frac {\nu^2}{8\pi^2}\int d^2 r \tilde{V}(r).
\end{equation}
We see immediately from this term that the symmetry-breaking
interactions favor equal population of the two layers, or in
pseudospin language they favor spin-textures where the pseudospin-orientation
is in the ${\bf\hat x}- {\bf\hat y}$ plane.

The right-hand side of Eq.(\ref{eq:sbeng}) can, after a straight-forward
but lengthy and tedious calculation, be expressed in
terms of the two-point correlations of $| \psi_0 \rangle$.
The calculations are similar to those in Ref.(\onlinecite{ahmz1}) and
involve commutators of magnetic translation operators and of
pseudospin operators.  The following identity enters the calculation
at several points:
\begin{equation}
[\overline S_{p}^{\mu},\, \overline S_{q}^{\nu}]=\frac {i}{\sqrt{A}}
\epsilon_{\lambda\mu\nu}\,
 \cos{\frac {\bf p \wedge q}{2}}\; e^{\frac {1}{2}{\bf p}
\cdot {\bf q}}\; \overline S_{p+q}^{\lambda}\, + \, \frac {i}{2\sqrt{A}}\,
\delta_{\mu\nu}\, \sin{\frac {\bf p \wedge q}{2}}\; e^{\frac {1}{2} {\bf p}
\cdot {\bf q}}\; \overline {\rho}_{p+q}.
\end{equation}
\noindent
In the limit of slowly varying spin-textures we obtain the following
result for the contribution of the symmetry-breaking term to the
energy of the spin-texture
\begin{equation}
E_{\rm sb}[{\bf m}]\simeq\int d^{2}r
\biggl\{ \beta_m (m^{z})^{2}\, +\frac {\rho_s^z}{2}(\nabla m^{z})^{2}-
\frac {\rho_s^z}{2}\bigl[(\nabla
m^{x})^{2}+(\nabla m^{y})^{2}\bigr] \biggr\},
\end{equation}
where
\begin{equation}
\rho_s^z={-\nu \over 32\pi^{2}}\int_{0}^{\infty}{dkV^{z}(k) h(k) k^{3}},
\label{eq:zstiffness}
\end{equation}
and $\beta_m = \beta_{H}+\beta_{xc}$ with
\begin{equation}
\beta_{xc}={\nu\over 8\pi^{2}}\int_{0}^{\infty}dk k V^{z}_k h(k).
\end{equation}
The total mass term is given by
\begin{equation}
\beta_m\equiv\beta_{H}+\beta_{xc}={-\nu \over 8\pi^{2}}\int_{0}^{\infty}dk\,
\bigl[ V^{z}(0)-V^{z}(k)\bigr] k\, h(k) .
\end{equation}
Notice that $V_A > V_E$ but that the intralayer interaction contains
an exchange term which reduces the effect of $V_A$.  Also
note that in the limit where the symmetry breaking interaction
is local, {\it i.e.}, the limit where $V^{z}_k$ is independent of
$k$, the exchange-correlation contribution to the coefficient of
the mass term vanishes.  We can understand this result in the
pseudospin language by noting that in this limit the symmetry
breaking interaction is proportional to $\sum_{i} (S_i^{z})^2$ which
(for spin-${1\over 2}$)
still commutes with all components of the total spin operator and
does not, despite appearances, destroy the SU(2) symmetry of the
Hamiltonian.  From another point of view we can understand this
result by noting that in $|\psi_0\rangle$ no two particles
can be at the same position and therefore they will not experience
a local interaction.   This property of $|\psi_0\rangle$
and the fact that $\beta_m$ vanishes for local $V^{z}_k$ can be confirmed
from the following identity satisfied by $h(k)$:
\begin{equation}
\int_{0}^{\infty} dk k h(k) =  - \nu.
\end{equation}
It is important to observe that in the limit of small layer
separations $V^{z}_k$ approaches $ \pi e^2 d$ which is local.
Because there is no contribution to $\beta_m$ from this local term
the mass coefficient ends up being proportional to the $d^2$
rather than proportional to $d$ at small $d$ as would be expected
based on naive considerations of the capacitance energy.

Including both SU(2) invariant contribution defined in
Eq.({\ref{nonlinearsigmamodel}) and the symmetry breaking contributions
found above,
the total energy-functional for a spin-texture is given by:
\begin{equation}
E_{\rm Tot}[{\bf m}]\simeq\int d^{2}r
\biggl\{ \beta_m (m^{z})^{2}\, +\frac {\rho_A}{2}(\nabla m^{z})^{2}+ \frac
{\rho_E}{2}\bigl[(\nabla
m^{x})^{2}+(\nabla m^{y})^{2}\bigr] \biggr\}
\label{eq:totaleng}
\end{equation}
where
\begin{equation}
\rho_A={-\nu \over 32\pi^{2}}\int_{0}^{\infty}{dkV^{A}_k h(k) k^{3}},
\label{eq:astiffness}
\end{equation}
and
\begin{equation}
\rho_E={-\nu \over 32\pi^{2}}\int_{0}^{\infty}{dkV^{E}_k h(k) k^{3}}
\label{eq:estiffness}
\end{equation}
This result is easy to understand.  The contribution to the
exchange-correlation energy which is dependent on the ${\bf \hat z}$
polarization of the pseudospin includes the Hartree energy
which favors $m_z=0$ and the exchange-correlation within the
layers.  The exchange correlation energy within each layer
increases superlinearly $\sim \rho^{3/2}$ with the layer density
so this term favors $m_z \ne 0$.  As discussed above the Hartree
energy is always larger for constant spin-densities.  The
term proportional to $(\nabla m^{z})^2$ in the energy density
captures the reduction of the
exchange-correlation energy from within each layer when the
density in the layer is not constant and therefore $\rho_A$ is dependent
only on the intra-layer interaction.  [$\rho_A = \rho_s^0$ at
all layer separations.  Because of the presence of the mass term,
this gradient term is not important at long wavelengths.]  On the other hand,
pseudospin-order in the ${\bf\hat x}-{\bf \hat y}$ plane represents interlayer
phase coherence.  As discussed earlier, an interlayer phase
relationship which changes as a function of position results
in a loss of interlayer correlation energy so that
$\rho_E$ depends only on inter-layer interactions.  In
Fig.[\ref{fig:massstiffnu1}] and Fig.[\ref{fig:massstiffnu3}]
we illustrate the dependence of
$\beta$ and $\rho_E$ on layer separation calculated
from the above expressions for $\nu=1$ and
$\nu = 1/3$ respectively.  We emphasize that these results are not expected to
be accurate at large layer separations.  We will compare
these results with estimates from exact diagonalization calculations in
Section \ref{sec:exactdiag}.

\subsection{Hartree-Fock Picture of Spin Textures and Gradient Expansion of
Energy Functional}

In this section we develop a Hartree-Fock picture to describe spin textures
and
derive the corresponding energy functional.  We also show that a gradient
expansion of this energy functional gives exactly the result we obtained in
the previous section.

We work in the Landau gauge: ${\bf A}=(0,\; Bx,\; 0)$.  The one-body
orbital
wave functions in the LLL in this gauge are:
\begin{equation}
\psi_X({\bf r})=\frac{1}{\sqrt{\pi^{1/2}L_y\ell}}\> e^{ik_yy}\>
e^{-\frac{(x-X)^2}{2\ell^2}},
\label{eq:c1}
\end{equation}
where $X=k_y\ell^2$ is the guiding center.

A particular class of single Slater determinants at $\nu =1$ in this gauge
can be written in the form
\begin{equation}
\vert\tilde{\psi}\rangle = \prod_X
\biggl(C_{X\uparrow}^\dagger\>\cos{\frac{\theta (X)}{2}} +
C_{X\downarrow}^\dagger\>\sin{\frac{\theta (X)}{2}}\> e^{i\varphi
(X)}\biggr)
\vert 0\rangle ,
\label{eq:c2}
\end{equation}
where $\vert 0\rangle$ is the fermion vacuum, $C_{X\uparrow
,\downarrow}^\dagger$ creates an electron in the upper (lower) layer in
orbit $\psi_X$ respectively.  In this state each
Landau gauge orbital is occupied by a single electron whose
pseudospin orientation is specified by the polar angles $\theta (X)$
and $\phi (X)$.  Each Landau gauge orbital is localized within
$\sim \ell$ of its guiding center.  We are interested in states
for which $\theta (X)$ and $\phi (X)$ vary slowly on the magnetic
length energy scale so that
$\vert\tilde{\psi}\rangle$ describes a spin texture in which
\begin{eqnarray}
m^z(x) &=&\cos{\theta(x)},\nonumber\\
m^x(x) &=&\sin{\theta(x)}\cos{\varphi(x)},\nonumber\\
m^y(x) &=&\sin{\theta(x)}\sin{\varphi(x)}.
\label{eq:c3}
\end{eqnarray}
$\vert\tilde{\psi}\rangle$ is not the most general spin texture,
because {\bf m} does not depend on $y$.  As a consequence
there is no spatial variation in charge density.

It is straightforward to evaluate the energy of $\vert\tilde{\psi}\rangle$.
For the following discussion we include the term in the Hamiltonian which
allows electrons to tunnel from layer to layer and whose consequences
will be explored in detail in a subsequent paper.\cite{II}
The Hamiltonian in this representation is
\begin{eqnarray}
\hat{H}&=&\hat{T} + \hat{V},\nonumber\\
\hat{T}&=&-t\> \sum_X \biggl( C_{X\uparrow}^\dagger C_{X\downarrow} +
C_{X\downarrow}^\dagger C_{X\uparrow}\biggr) ,\nonumber\\
\hat{V}&=&\frac{1}{2}\> \sum_{{X_1X_2X_3X_4}\atop
{\sigma_1\sigma_2}}\>
V_{X_1X_2X_3X_4}^{\sigma_1\sigma_2}\> C_{X_1\sigma_1}^\dagger\;
C_{X_2\sigma_2}^\dagger\; C_{X_4\sigma_2}\; C_{X_3\sigma_1} ,
\label{eq:c4}
\end{eqnarray}
where $\sigma =\uparrow ,\downarrow$ is the layer index,
\begin{eqnarray}
V^{\uparrow\uparrow} &=& V^{\downarrow\downarrow} = V^A\qquad ,\qquad
V^{\uparrow\downarrow} = V^{\downarrow\uparrow} = V^E,\nonumber\\
V_{X_1X_2X_3X_4} &=& \int d{\bf r}_1d{\bf r}_2\> V({\bf r}_1 - {\bf r}_2)
\psi_{X_1}^\ast ({\bf r}_1) \psi_{X_2}^\ast ({\bf r}_2) \psi_{X_3}({\bf
r}_1)
\psi_{X_4}({\bf r}_2).
\label{eq:c5}
\end{eqnarray}
Hence we have
\begin{eqnarray}
\langle\tilde{\psi}\vert\hat{T}\vert\tilde{\psi}\rangle &=& -2t\> \sum_X
\cos{\frac{\theta (X)}{2}}\; \sin{\frac{\theta (X)}{2}}\; \cos{\varphi
(X)}\\
&=& -t\> \sum_X m^x (X)\\
&=& -\frac{t L_y}{2\pi\ell^2}\> \int dx\> m^x (x)\nonumber\\
&=& -\frac{t}{2\pi\ell^2}\> \int d^2{\bf r}\> m^x ({\bf r}).
\label{eq:c6}
\end{eqnarray}
In the last step we allow the possibility
that $m^x$ depends on $y$ as well.  We see this result
agrees exactly with the tunneling energy we obtained previously.  The
contribution
from interactions can also be evaluated using Wick's theorem:
\begin{eqnarray}
\langle\tilde{\psi}\vert\hat{V}\vert\tilde{\psi}\rangle
&=& \frac{1}{2}\>
\sum_{{X_1X_2X_3X_4}\atop {\sigma_1\sigma_2}}
V_{X_1X_2X_3X_4}^{\sigma_1\sigma_2} \biggl(\langle C_{X_1\sigma_1}^\dagger
C_{X_3\sigma_1}\rangle\langle C_{X_2\sigma_2}^\dagger
C_{X_4\sigma_2}\rangle -
\langle C_{X_1\sigma_1}^\dagger C_{X_4\sigma_2}\rangle\langle
C_{X_2\sigma_2}^\dagger C_{X_3\sigma_1}\rangle\biggr)\nonumber\\
&=& \frac{1}{4}\> \sum_{X_1X_2} \Biggl\lbrace -E^A(X_1-X_2)\nonumber\\
&+& [D^A(X_1-X_2) -
D^E(X_1-X_2) - E^A(X_1-X_2) ]\> m^z (X_1) m^z (X_2) \nonumber\\
&&-  E^E(X_1-X_2) [m^x (X_1) m^x (X_2) + m^y (X_1) m^y
(X_2)]\Biggr\rbrace \label{eq:c7}
\end{eqnarray}
In this equation we have absorbed the Hartree energy of the
system for equal layer densities into the zero of energy.
The quantities $D(X) \equiv V_{X+Y,Y,X+Y,Y}$ and $E(X) \equiv V_{X+Y,Y,Y,X+Y}$
are the direct and exchange two-body integrals for both intr{\it a} and
int{\it er} layer interactions.   The above equation has clear physical
content.  The first term in the final form of the
equation is the exchange energy in the absence of
pseudospin polarization.  The second term is the
Hartree charging energy including a exchange correction.
The third term is the exchange energy due to interlayer
coherence, which is the source of the loss of exchange energy when $m^x$
and $m^y$ change.  We now make a gradient expansion by writing
\begin{eqnarray}
m^\nu (x_1) m^\nu (x_2) &=& [m^\nu (x_1)]^2 + m^\nu (x_1)(x_2 -
x_1)\biggl(\frac{\partial}{\partial x} m^\nu (x_1)\biggr)\nonumber\\
&&+ \frac{1}{2}\> m^\nu (x_1)
(x_2 - x_1)^2\> \biggl(\frac{\partial^2}{\partial x^2}
m^\nu
(x_1)\biggr) +\cdots,
\label{eq:c8}
\end{eqnarray}
Replacing the summation over guiding centers by integrations,
we easily obtain recover Eq.[~\ref{eq:totaleng}].  (The coefficients
of the gradient terms are proportional to the second moments
of the exchange integrals, $\sum_X X^2 E(X)$.) In this formulation
we see explicitly that the leading gradient corrections are
adequate as long as the pseudospin orientation changes slowly on
the scale of $\ell$.

\subsection{Hubbard-Stratonovich transformation approach to the effective
action}

In subsection B above, we
derived the effective action by calculating
the energy functional for spin textures.  The dynamical term
in the effective action was determined by requiring that
the Lagrangian implies the correct equation of motion for the
spin textures.  In this section we briefly sketch an alternate route for
deriving the same effective action.
The basic idea is the familiar Hubbard-Stratonovich (HS) transformation.
We introduce auxiliary fields to decouple
the interaction and integrate out the fermionic degrees of freedom to
obtain the effective action of the auxiliary fields. The auxiliary fields are
essentially the order parameters.

It is inconvenient to project onto the lowest Landau level until the
end of the calculation so we work with the full Hamiltonian which
has the form
\begin{equation}
\hat{H}=\hat{H}_0+\hat{V},
\end{equation}
where
\begin{equation}
\hat{H}_0=\sum_{j}{1\over 2m}
\left[{\bf p}_j-{e\over c}{\bf A}({\bf r}_j)\right]^2
\end{equation}
is the kinetic energy of the particles in the presence of the external
magnetic field.  This approach is more easily implemented if we have
a delta-function-like interaction $V({\bf r})=V_0\delta({\bf r})$ and
so we discuss this case first:
\begin{equation}
\hat{V}={V_0\over 2}\sum_{q}{\rho_q\rho_{-q}
=-V_0\sum_{q}{\bf S}_q\cdot{\bf S}_{-q}
-{V_0\over 4}\sum_{q}\rho_q\rho_{-q}}.
\end{equation}
The advantage of a delta function interaction is that it can be expressed
in terms of spin operators,\cite{hsnote1} which makes it possible to
decouple the interaction in terms of spin auxiliary fields. Written
in the above particular form, we will find that the saddle point of the
auxiliary fields corresponds to the Hartree-Fock mean-field
Hamiltonian.\cite{schulz}
The partition function is\cite{hsnote2}
\begin{equation}
Z(\beta)={\rm Tr}\left\{e^{-\beta \hat{H}_{0}}\hat{S}(\beta)\right\},
\end{equation}
where
\begin{equation}
\hat{S}(\beta)=T_{\tau}e^{-\int_{0}^{\beta}{\hat{V}(\tau)}},
\end{equation}
and
\begin{equation}
\hat{V}(\tau)=e^{\hat{H}_0\tau}\hat{V} e^{-\hat{H}_0\tau}.
\end{equation}
$T_{\tau}$ is the (imaginary) time ordering operator.
Now we can introduce a vector auxiliary field ${\bf h}(\tau,{\bf r})$ and
a scalar auxiliary field $\phi(\tau,{\bf r})$ to
 decouple $V$:

\begin{equation}
Z(\beta)=\int{D{\bf h}}D\phi
e^{-\int_{0}^{\beta}\int{d^2 r}{d\tau} ({1\over 4V_0}{\bf h}^2+{1\over V_0}
\phi^2)}
{\rm Tr}
\left\{e^{-\beta \hat{H}_0}T_{\tau}
e^{-\int_{0}^{\beta}{d\tau}\int{d^2 r}(-{\bf S}\cdot
{\bf h}-\rho\phi)}\right\}.
\end{equation}

After the HS decoupling, we find the fermionic Hamiltonian becomes quadratic
so we can (at least in principle),
carry out the trace over the fermion degrees of freedom and hence obtain
the effective action in terms of the auxiliary fields ${\bf h}$ and $\phi$.
In doing that, however, we still need to make approximations. We notice that
the direction fluctuations of ${\bf h}$ are massless (due to
broken $SU(2)$
symmetry) while the fluctuations of $\phi$ are massive.\cite{schulz}
It is therefore
a good approximation to assume that $\phi$ is ``frozen"
to be a constant in space and time so it only
contributes a chemical potential like term in the fermion Hamiltonian and
is hence unimportant. Thus
 we will concentrate on the fluctuations of ${\bf h}$.

In computing the trace we notice that the trace is over
nothing but the propagator
of a system governed by the time dependent Hamiltonian
\begin{equation}
\hat{H}(\tau)=\hat{H}_0-\int{d^2 r}\,
{\bf S}({\bf r})\cdot{\bf h}(\tau,{\bf r}) +{\rm const}.
\end{equation}
in imaginary time. Since there is always a large one body gap for
$\hat{H}(\tau)$ while the collective modes of ${\bf h}$ are gapless (which
means $\hat{H}(\tau)$ is slowly varying),
we can use the adiabatic (or Born-Oppenheimer) approximation to
evaluate the propagator\cite{stone}:
\begin{equation}
{\rm Tr}
\left\{e^{-\beta\hat{H}(\tau)}\right\}
=e^{i\gamma[\Gamma]-\int_{0}^{\beta}{E(\tau)d\tau}},
\end{equation}
where $\gamma[\Gamma]$ is the Berry's phase determined by\cite{stone}

\begin{equation}
i{d\gamma\over d\tau}=-\langle\Psi(\tau)|{d\over d\tau}\Psi(\tau)\rangle,
\end{equation}
where $|\Psi(\tau)\rangle$ is the ground state
of $H(\tau)$ and $E(\tau)$ is the
ground state energy.
In order to proceed, we need to find $|\Psi(\tau)\rangle$ and $E(\tau)$.
It is now possible to take the strong magnetic field limit.
It is both plausible and easily checked that in this case
$|\Psi(\tau)\rangle$ is nothing but our familiar spin texture state:

\begin{equation}
|\Psi(\tau)\rangle=e^{i\overline{O(\tau)}}|\Psi_0\rangle,
\end{equation}
where $|\Psi_0\rangle$ is the fully polarized state in $\hat{z}$ direction and

\begin{equation}
O(\tau)=\sum_{j}{{\bf S}_j\cdot{\bf \Omega}(\tau, {\bf r_j})},
\end{equation}
where ${\bf \Omega}(\tau, {\bf r})={\bf\hat{z}}
\times{\bf m}(\tau, {\bf r})$ and
${\bf m}(\tau, {\bf r})$ is a unit vector in the direction of ${\bf h}$.

Now let us look at the Berry's phase term:
\begin{eqnarray}
id\gamma &=& 1-\langle\Psi(\tau)|\Psi(\tau+d\tau)\rangle\nonumber\\
        &=&
1-\langle\Psi(\tau)|e^{i\overline{O}(\tau+d\tau)}e^{-i\overline{O}(\tau)}
|\Psi\rangle(\tau)\nonumber\\
        &=&
1-\langle\Psi(\tau)|e^{i\partial_{\tau}\overline{O}(\tau)d\tau-{1\over 2}
[\overline{O}(\tau), \partial_{\tau}\overline{O}(\tau)]d\tau}|\Psi(\tau)\rangle
\nonumber\\
        &=& -i\langle\Psi(\tau)|\partial_{\tau}\overline{O}(\tau)+{i\over 2}
[\overline{O}(\tau),
\partial_{\tau}\overline{O}(\tau)]|\Psi(\tau)\rangle d\tau\nonumber\\
        &=& i{\rho\over 2}d\tau\int{d^2{\bf r}}\,{\bf A}({\bf m}
(\tau, {\bf r}))\cdot\partial_{\tau}{\bf m}(\tau, {\bf r}).
\end{eqnarray}
Here $\rho=1/(2\pi \ell^2)$ is the density of electrons. Hence we find that
the Berry's phase term is exactly the dynamical (time dependent) term we
obtained previously:

\begin{equation}
i\gamma=i{\rho\over 2}\int{d\tau}\int{d^2{\bf r}}\,{\bf A}({\bf m}
(\tau, {\bf r})
)\cdot\partial_{\tau}{\bf m}(\tau, {\bf r}).
\end{equation}

Now let us calculate $E(\tau)$:
\begin{eqnarray}
E(\tau) &=& \langle\Psi(\tau)|H(\tau)|\Psi(\tau)\rangle=\langle\Psi_0|
e^{-i\overline{O}(\tau)}H(\tau)e^{i\overline{O}(\tau)}|\Psi_0\rangle\nonumber\\
&=& \langle\Psi_0|H-i[\overline{O},H]-{1\over 2}[\overline{O},[\overline{O},H]]
+\cdots|\Psi_0\rangle\nonumber\\
&=& -{\rho\over 2}\int{d^2{\bf r}}\,{\bf h}\cdot{\bf m}+{1\over 8}\rho \ell^2
\int{d^2{\bf r}}\,(\nabla h^{\mu})\cdot(\nabla m^{\mu})+\cdots.
\end{eqnarray}
The first term is just the gain of energy
by polarizing the spins in the direction of
the external field. The second term comes from the nonlocality induced
by the projection to the LLL, so the spins cannot take full advantage of the
external field if its direction is changing in space. This is exactly the
physics that is responsible for the stiffness. Hence the effective action is

\begin{equation}
S_E[{\bf \phi}]=i\gamma[\Gamma]+\int_{0}^{\beta}{d\tau}\int{d^2 r}\,\{{V_0\over
4}h^2-{\rho\over 2}h+{1\over 8}\rho\ell^2(\nabla h^{\mu})
\cdot(\nabla m^{\mu})\}.
\end{equation}
Here $h=\vert {\bf h}\vert$. We find the action has its minimum when

\begin{equation}
{\bf h}=\rho V_0{\bf\hat{n}},
\end{equation}
where ${\bf\hat{n}}$ is a constant unit vector. At this point the one body
Hamiltonian given by ${\bf h}$ is exactly the Hartree-Fock Hamiltonian.
The fluctuations of the magnitude
of ${\bf h}$ are
massive and we can integrate them out to obtain the non-linear
sigma model:

\begin{equation}
S_E[{\bf m}]=i\gamma[\Gamma]+{1\over 2}\rho_s\int_0^{\beta}{d\tau}\int{d^2r}
\, (\nabla{\bf m})^2,
\end{equation}
where the stiffness $\rho_s=\rho^2 \ell^2 V_0/4=V_0/(16\pi^2 \ell^2)$,
is exactly what we obtained from the spin texture calculations.

The above discussion was for the case of short range interactions.
To generalize to the finite-range-interaction case
we separate a $\delta$ function like part from the interaction, follow
the same procedures as illustrated above, and treat the remaining part of the
interaction as a perturbation.  Note that non-delta function like
interactions are not necessarily $SU(2)$ invariant.
At the first order in this perturbation theory (where
$\vert\Psi(\tau)\rangle$ is unaffected), we recover all the results obtained
in previous sections.

\section{Charged Objects in the System}
\label{sec:vortices}
\subsection{Skyrmions}

We start by briefly reviewing and commenting on results for charged
excitations obtained by Sondhi {\it et al.}\cite{sondhi} for
a single-layer system of spin-$1/2$ particles with no Zeeman coupling.
These results can be directly taken over to the
double-layer system with $d/\ell=0$ (which is therefore $SU(2)$ invariant).
In this case
one finds\cite{kallin,rasolt}, as discussed in the following section and
mentioned previously, that the double-layer
system with spontaneous interlayer phase coherence supports
$neutral$ gapless `spin-wave' excitations which disperse quadratically
in the long wavelength limit.
This property of spin waves is characteristic of
isotropic ferromagnetism.  However, in contrast to the case of
quantum Heisenberg ferromagnets on a lattice, quantum Hall systems
also possess charge degrees of freedom and are more
analogous to itinerant electron ferromagnets.  For example,
after a spin is flipped, it can be scattered to other orbital states
and carry charge throughout the system.
In addition to the gapless spin wave excitations,
there are $gapful$ charged excitations.  Only charged excitations
can contribute to the conductivity so that the low temperature
transport coefficients will be activated with an activation
energy which is half the charge gap.  Some of the low-energy
charged excitations can be generated from the topologically non-trivial spin
configurations discussed below.

We have seen in Section III that the physical charge density of the system
at $\nu = 1/m$ is $1/m$ of the topological charge density.
As a consequence of this the
topological solitons, {\it i.e.}, skyrmions,
carry $1/m$ units of physical charge
(see Fig. [\ref{fig:fig41}]).\cite{leekane,sondhi}
Inside a skyrmion the spins wrap around the unit order-parameter
sphere exactly once.\cite{rajaraman} In the $SU(2)$ invariant
case with Coulomb interactions, Sondhi {\it et. al.} have shown
that skyrmions are the lowest energy
charged excitations of the system at $\nu=1$.\cite{sondhi}
This is not surprising
because for the skyrmion spin configuration, the spins are nearly
parallel locally, so the exchange energy is only slightly reduced.
In contrast, for ordinary single particle excitations
[see Fig.(\ref{fig:fig42})],
an added electron has its spin opposite to the others and has
no exchange energy.  As pointed out earlier, in the SU(2) invariant limit
we know the {\it exact\/} spin
stiffness. Hence the exact energy of a single (large scale)
skyrmion can be obtained\cite{rajaraman,sondhi}:
\begin{equation}
E_{s}=4\pi\rho_s.
\end{equation}
For the case of a system with Coulomb interactions at $\nu=1$,
we obtain from the non-linear sigma model energy expression
\begin{equation}
E_s={1\over 4}\sqrt{{\pi\over 2}}{e^2\over \epsilon \ell}.
\end{equation}

It is important to realize that since the total particle number is fixed
in our derivation of the non-linear sigma model energy expression,
$E_s$ actually gives the energy to introduce an isolated
skyrmion or anti-skyrmion into the bulk of the system {\it and}
maintain charge neutrality
by adding or subtracting charge from  the edge of the system.
(For related careful discussions of quasiparticle energies in the
fractional quantum Hall effect see
Refs.[\onlinecite{quasimorf,quasimacd}].)
This energy must be subtracted off if we wish to calculate
$ \epsilon_{\pm}$, the energy to add ($+$)or subtract ($-$)
$1/m$ electrons from the system at $\nu = 1/m$ in the form of a
skyrmion or anti-skyrmion spin-texture.  It follows\cite{quasimacd} that
\begin{equation}
\epsilon_{\pm} = \pm \nu \xi(\nu) + E_s
\label{qpenergy}
\end{equation}
where $\xi(\nu)$ is the energy per electron in the incompressible
ground state at $\nu=1/m$.  (For the Coulomb interaction case
$\xi(1)= - \sqrt{\pi/8} (e^2/\ell) \approx -0.6266 (e^2 / \ell)$
and\cite{engmacd} $\xi(1/3) \approx -0.4100 (e^2/ \ell)$.)  The chemical
potential for $\nu > 1/m $ is $ m \epsilon_{+}$ while the
chemical potential for $\nu < 1/m$ is $ -m \epsilon_{-}$.
For $\nu =1$ it follows from the above that $\epsilon_{+} =
- (1/4)  \sqrt{\pi/2} (e^2 / \ell) $ and
$\epsilon_{-} = (3/4) \sqrt{\pi/2} (e^2 / \ell) $.  The energies
of the localized quasiparticle excitations of Fig.[\ref{fig:fig42}]
cannot be reliably calculated from the non-linear sigma model energy
expression but for $\nu =1 $ the microscopic energy
 can be calculated analytically.
For these excitations $\epsilon_{+} = 0$ (the spin-reversed added
electron has {no} exchange energy!) and $\epsilon_{-}
= \sqrt{\pi/2} (e^2/ \ell)$.  In the absence of Zeeman coupling
it follows that the lowest energy particles and holes are
both formed from skyrmion spin textures.  The lowest energy
particle-hole excitation is a skyrmion anti-skyrmion pair which
has energy $2 E_s$.   This is only
one half of that of the ordinary particle-hole pair in the case of
the Coulomb interaction.\cite{sondhi}  These results
receive unequivocal support from numerical calculations.\cite{rezayi}
Using our result for the spin-stiffness we can extend this
analysis to the case of $\nu = 1/3$.  At this filling
factor $E_s \approx 0.0116 (e^2 / \ell)$, $ \epsilon_{+} \approx
- 0.1251 (e^2 / \ell)$,  $\epsilon_{-} \approx 0.1483 (e^2 / \ell)$.
In the limit of large Zeeman energy at this filling factor both
the localized quasihole and quasiparticle excitations will be
completely spin-polarized.  The quasiparticle and quasihole energies
in this limit have been estimated\cite{quasimacd,quasimorf}
to have the values $\epsilon_{+} \approx -0.120 (e^2 / \ell)$ and
$\epsilon_{-} = 0.2337$.  Again the skyrmion and antiskyrmion states
possible at zero Zeeman coupling have lower energy, although only
barely so in the quasiparticle case.  The particle-hole creation
energy $2 E_s$ for $\nu = 1/3$ is approximately four times larger
than in the large Zeeman coupling limit.  This is
consistent with results from the finite size exact-diagonalization
study of Chakraborty {\it et. al.} who found\cite{chakra} that
in the absence of Zeeman coupling, quasiparticle energies at $\nu=1/3$
could be reduced by flipping a single spin.

Another consequence of skyrmions being the lowest energy
charged excitation is
that in
finite size systems on a sphere, the total spin of the ground state changes
suddenly from ${N\over 2}$ to zero or one half (for odd and even $N$
respectively) when one changes the particle number from $N$ to $N\pm 1$,
where $N$ is the Landau level degeneracy without spin.\cite{rezayi,sondhi}
(We have verified that the same property holds for $\nu =1/3$
in agreement with the analysis of the preceding paragraph.)
This is because
when a skyrmion is put on a sphere, the spin configuration is like
a hedgehog and it is plausible that the total spin is essentially
zero.\cite{sondhi}  [As we discuss below the {\it total\/} angular momentum
$\bf J$ is precisely zero.]
Note however that the system is still ferromagnetic in the
sense that it is not a {\em local\/} singlet\cite{girvinappendix} and
its Zeeman susceptibility still diverges.

In this case the topological charge density is uniformly distributed
on the sphere and the skyrmion is unfrustrated. The situation
is very different
for a system with rectangular geometry with periodic boundary conditions,
i.e, the geometry of a torus. In this case the skyrmion is frustrated and
hence squeezed by the finite size effect, so its size is much smaller than
the system size. This can be understood by looking at the effect of
periodic boundary conditions on the energy of a skyrmion.
We know the gradient energy term
is scale invariant, and for the ideal skyrmion solution,
the energy is minimized to be
$4\pi\rho_s$.  However an ideal solution does not satisfy the
boundary conditions.  In a rectangle a skyrmion has to be distorted
near the boundary so the energy from the stiffness term will increase.
It is obvious that the smaller the size of the skyrmion, the
smaller the energy cost due to the boundary effect will be.
However the
skyrmion cannot be too small because it costs too much Coulomb energy (which
wants the skyrmion to be as large as possible so that the excess charge
will be distributed as uniformly as possible).  The optimal size of a
skyrmion on a rectangle is determined by a competition between
stiffness and Coulomb energies and
as a consequence the total spin $S$ of the ground
state of the $N+1$ particle system is size dependent.

We can estimate the difference between $(N+1)/ 2$ and $S$
(i.e., the number of spins flipped in the ground state)
in the following way.  Keeping the two leading terms, the
energy of a spin texture at $\nu = 1$ is
\begin{equation}
E={1\over 2}\rho_s\int{d^2{\bf r}\,(\nabla{\bf m})^2}+{1\over 2}
\int{d^2{\bf r} d^2{\bf r}'}\,V({\bf r}-{\bf r}')\rho({\bf r})\rho({\bf r}'),
\label{eq:skeng}
\end{equation}
where $\rho$ is the topological charge density and $V$ is the Coulomb
interaction. The energy of a skyrmion with
linear size $\lambda \gg \ell $ in  a system with linear size $R$ is
\begin{equation}
E(\lambda)=4\pi\rho_s+A({\lambda\over R})^2+B{\ell
\over \lambda},
\label{eq:skyrm1234}
\end{equation}
where $A$ and $B$ are positive constants with units of energy. The first term
in Eq.(\ref{eq:skyrm1234})
is the usual energy of an unfrustrated, infinite size skyrmion, and
the last two are finite size corrections from the two terms in
Eq.[\ref{eq:skeng}].
respectively.  Minimizing $E$ with respect to $\lambda$ gives
\begin{equation}
\lambda\propto R^{2\over 3}\ell^{1\over 3}.
\end{equation}
The number of spins flipped in a skyrmion with size $\lambda$ and long distance
cutoff $R$ is\cite{sondhi}
\begin{equation}
\triangle S\propto\lambda^2\ln({R\over \lambda})\propto N^{2\over 3}\ln N,
\label{eq:fspin}
\end{equation}
where we used the fact that $N\propto R^2$. From Eq.[\ref{eq:fspin}]
we see that as $N$ becomes
large, the number of flipped spins in the optimal skyrmion state
gets large but is always small compared to $N$
so that the ground state is almost fully polarized.

Skyrmion spin-textures produce
an excitation energy which is independent of the texture size
as long as the size is large compare to microscopic lengths
$(\ell)$ and small compared to the system size.  This leads to
dramatic finite-size effects which are typified by the qualitative differences
between ground state spin quantum numbers for electrons in a rectangle
and on a sphere, where the state is
unpolarized.  For the case of a square with periodic boundary conditions,
a crude estimate gives the coefficient in front of $N^{2\over 3}\ln N$ in
Eq.(\ref{eq:fspin}) to be 0.26.
In Fig.(\ref{fig:sqspin}) we compare this estimate
with results from finite-size exact diagonalization calculations for
square boundary conditions\cite{haldanetorus} and find qualitative agreement.
Using the composite fermion theory,
Jain and Wu\cite{jain} gave an alternative explanation of the fact that
the total spin goes to zero for an $N+1$ electron system on a sphere.
Their theory does not distinguish between the sphere and torus and we do not
believe the results on a torus can be easily understood in their formalism.
The quantitative agreement with the ground state energy on the sphere
and the qualitative agreement for the polarization on the torus lend
strong weight to the skyrmion picture.

We close this discussion by noting that it is possible to write down
simple microscopic variational wave functions for the skyrmion, both
in the plane and on the sphere.  Consider the following state in the plane
\begin{equation}
\psi_\lambda = \prod_m \left(\begin{array}{c} z_m \\ \lambda \\ \end{array}
\right)_m \Psi_{V},
\label{eq:skyrmicro}
\end{equation}
where $\Psi_{V}$ is defined in Eq.(\ref{eq:halperin111}),
$()_m$ refers to the spinor for the $m$th particle, and $\lambda$ is
a fixed length scale.  This is a skyrmion because
it  has its spin purely down at the origin
(where $z_m = 0$) and has spin purely up at infinity (where $z_m \gg
\lambda$).  The parameter $\lambda$ is simply the size scale of the
skyrmion\cite{sondhi,rajaraman}.   Notice that in the limit $\lambda
\longrightarrow 0$ (where the continuum effective action is invalid, but
this microscopic wave function is still sensible) we recover a fully
spin polarized filled Landau level with a charge-1 Laughlin quasihole
at the origin.  Hence the number of flipped spins
interpolates continuously from zero to infinity as $\lambda$ increases.

In order to analyze the skyrmion wave function in Eq.(\ref{eq:skyrmicro}),
 we use the Laughlin plasma analogy.
In this analogy the norm of $\psi_\lambda$,
 $Tr_{\{\sigma\}}\int D[z] |\Psi[z]|^2$
is viewed as the partition function of a Coulomb gas.
In order to  compute the density distribution
we simply need to take a trace over the spin
\begin{equation}
Z=\int D[z]\, e^{2\left\{\sum_{i>j} \log|z_i
-z_j| + \frac{1}{2}\sum_k \log(|z_k|^2+
\lambda^2) - \frac {1}{4}  \sum_k |z_k|^2\right\}}.
\label{eqm20}
\end{equation}
This partition function describes the usual logarithmically interacting Coulomb
gas with uniform background charge plus
a spatially varying impurity back ground charge $\Delta\rho_b(r)$,
\begin{equation}
\Delta\rho_b(r)\equiv-\frac {1}{2\pi} \nabla^2 V(r)=
-\frac{\lambda^2}{\pi(r^2+\lambda^2)^2},
\label{eqm30}
\end{equation}
\begin{equation}
V(r)=\frac{1}{2}\log(r^2+\lambda^2).
\label{eqm40}
\end{equation}

For large enough scale size $\lambda \gg \ell$, local neutrality of the
plasma\cite{jasonho}
implies that the excess electron number density is precisely
$\Delta\rho_b(r)$, so that Eq.(\ref{eqm40})
is in agreement with the standard result for the
topological density.\cite{rajaraman}

For a complete microscopic analytic solution valid for arbitrary
$\lambda$, we use the fact that the proposed
manybody wave function is nothing but a Slater determinant of the
single particle states $\phi_m(z)$,
\begin{equation}
\phi_m(z)=\frac {z^m}{\sqrt{2\pi 2^{m+1} m! (m+1+\frac {\lambda^2}{2})}}
{z\choose \lambda} e^{-\frac {|z^2|}{4}}.
\label{eqm50}
\end{equation}
The electron number density is then
\begin{equation}
\Delta n^{(1)}(z)\equiv\sum_{m=0}^{N-1}
|\phi_m(z)|^2 -\frac {1}{2\pi},
\label{eqm60}
\end{equation}
which yields
\begin{equation}
\Delta n^{(1)}(z)=\frac {1}{2\pi} \biggl( \frac {1}{2} \int_0^1 d\alpha
\, \alpha^{\frac {\lambda^2}{2}} e^{-\frac {|z|^2}{2} (1- \alpha)}
(|z|^2+\lambda^2) -1 \biggr).
\label{eqm70}
\end{equation}
Similarly, the spin density distribution $S^z(r)$ can
be obtained, and it also agrees with that for the standard skyrmion in
\cite{rajaraman} the large $\lambda$ limit.
We have also computed the
skyrmion creation energy from the spin-dependent
pair correlation functions of the plasma
following the same procedure as in Ref.(~\onlinecite{quasimacd}).
Fig.(\ref{fig:microskr}) shows a plot of this energy as a function
of scale size $\lambda$ and shows that the microscopic formula
gives the correct asymptotic value of one-half
the quasi-hole energy for the large $\lambda$-limit,
in which the continuum field theoretic picture holds exactly.
As the core size decreases, the skyrmion energy increases due to the
increasing Coulomb charging energy.   However it does not diverge as the
naive extrapolation of the field theoretic expression would.

Finally, we note that by replacing $z\choose \lambda$ by
$z^n\choose \lambda^n$, we can generate a skyrmion with a Pontryagin
index $n$.

The skyrmion wave function has a particularly simple form on a sphere.
On a sphere with radius $R=S^{1/2}\ell$ where $S$ is an integer or a half
integer, the number of a flux quanta is $N_s=2S$.  The single particle kinetic
energy is\cite{fdmhz1}
\begin{equation}
T=\frac{1}{2}\omega_c\vert {\bf\Lambda}\vert^2 /S,
\end{equation}
where
\begin{equation}
{\bf\Lambda} = {\bf r}\times \bigl(-i\nabla + e{\bf A}({\bf r})\bigr)
\end{equation}
is the kinetic angular momentum.  One can show that ${\bf L}={\bf\Lambda} +
S {\bf\Omega}$ (where ${\bf\Omega}=\frac{{\bf r}}{\vert {\bf
r}\vert}$)
is the generator of rotations for the system, i.e.:
\begin{equation}
[L^\alpha , X^\beta ] = i\epsilon_{\alpha\beta\gamma} X^{\gamma} ,
\end{equation}
where ${\bf X}$ is any vector.  We also have
\begin{equation}
\vert {\bf\Lambda}\vert^2 = \vert {\bf L}\vert^2 - S^2,
\end{equation}
so the eigenvalues of $\vert {\bf\Lambda}\vert^2$ have the form
$(n+S)(n+S+1)-S^2$, where $n$ is an integer.  For $n=0$, one obtains the
LLL
energy $T=\frac{1}{2}\omega_c$, and the degeneracy is $2S+1=N_s+1$.  If we
use
the Dirac gauge ${\bf A} = \frac{S}{eR}{\bf\hat{\phi}}
\cot\theta$, everything can
be
easily expressed in terms of spinor coordinates:
$u=\cos{\frac{\theta}{2}}\>
\exp(i\frac{\phi}{2})$, $v=\sin{\frac{\theta}{2}}\>
\exp(-i\frac{\phi}{2})$.
In this representation\cite{fdmhz1}
\begin{eqnarray}
L^+ &=& u\partial_v ,\\
L^- &=& v\partial_u ,\\
L^z &=& \frac{1}{2}\> ( u\partial_u -
v\partial_v) .
\end{eqnarray}
The LLL wave functions are simply homogeneous polynomials of $u$ and $v$
of degree $2S$.  The filled LLL single Slater determinant is just
\begin{equation}
\prod_{i<j}^N (u_i v_j - u_j v_i),
\end{equation}
where $N=N_s+1$.  The single antiskyrmion (that carries charge -1)
wave function is simply
\begin{equation}
\psi_{as}=\prod_{k=1}^{N-1} \left(\begin{array}{c}
v_k\\
-u_k\end{array}\right) \prod_{i<j}^{N-1} (u_iv_j - u_jv_i),
\end{equation}
where we have explicitly put in the fact that the total number of
particles is now $N-1$.
The spin configuration of $\psi_{as}$ is that of a hedgehog
(with spins pointing inside toward the center of the sphere) since
the ratio of $|u_k|$ to $|v_k|$ varies as $\cot(\theta/2)$.  This state is
neither
an eigenstate of ${\bf S}_{\rm tot}$, nor an eigenstate of ${\bf L}_{\rm
tot}$.  It is however a singlet of the total angular momentum ${\bf J}$:
\begin{equation}
{\bf J}={\bf L}_{\rm tot} + {\bf S}_{\rm tot}.
\end{equation}
Physically this means the state $\vert\psi_{as}\rangle$ is invariant
under a spin rotation followed by an identical space rotation, which is
clear from the uniform nature of the hedgehog
spin configuration.

If we project $\psi_{as}$ onto the subspace of $S_{\rm tot}=0$ (or ${1\over 2}$
if we have an odd number of particles), we
automatically get $L_{\rm tot} = 0$ (or ${1\over 2}$).
So the projected state will be invariant
under
both spin and space rotation.  This state should have good overlap with the
exact ground state.

The skyrmion (that carries charge $+1$) wave function has a similar form:
\begin{equation}
\psi_{s}=\prod_{k=1}^{N+1} \left(\begin{array}{c}
\partial_{u_k}\\
\partial_{v_k}\end{array}\right)
\prod_{i<j}^{N+1} (u_iv_j - u_jv_i) .
\end{equation}
The spin configuration of this state is exactly the opposite of $\psi_{as}$,
i.e., it is like a hedgehog with all spins pointing outward.

\subsection{Merons}

When $d/ \ell \ne 0$, the ${\bf\hat{z}}$ component of the order parameter is
massive and the system has $U(1)$ symmetry.
In this case, there is another class of topologically stable
charged objects,  merons.\cite{gross,affleck,early}
As shown in the following, merons [see Fig.(\ref{fig:meron})]
carry one half unit of topological charge and hence\cite{tcrmk} $1/2m$ units of
electron charge.  Far away from the core of a meron
the order parameter lies in the (massless) XY plane and forms a
vortex configuration with $\pm$ vorticity, while inside the core region
the order parameter smoothly rotates either up or down out of the XY plane.
Hence there are four flavors of merons.
The energy of a single meron diverges
logarithmically with the system size with a coefficient
proportional to the inter-layer spin stiffness.
The interaction between merons has a contribution
from the stiffness energy which is also logarithmic,
attractive for opposite vorticity pairs and repulsive for same
vorticity pairs.  These properties are exactly the same as the vortices
in the classical XY model.
In order to determine the sign of the charge carried by a meron, one has to
specify both its vorticity and the spin configuration in the core region.
Merons will also have a long range $1/r$ interaction due to their
charges which is attractive for oppositely charged merons and
repulsive for like-charged merons.

The fact that merons carry topological charge one half can
be seen by the following argument.  Imagine
a vortex in the spin system.  If an electron circles the vortex at a large
distance, its spin rotates through $2\pi$.  This induces a Berry's phase
of $\exp(i2\pi S) = -1$ which is equivalent to that induced by a charge moving
around one-half of a flux quantum.  Since $\sigma_{xy} = e^2/mh$, the vortex
picks up charge $1/2m$.  The topological charge of a meron
can also be understood by considering a variational function for the
meron spin texture:
\begin{equation}
{\bf m}=\left\{\,\sqrt{1-(m^z(r))^2} \cos\varphi\, ,
\sqrt{1-(m^z(r))^2} \sin\varphi\, , m^z(r)\right\}.
\end{equation}
The local topological charge density
calculated from $\delta\rho= -{1\over 8\pi}\epsilon_{ij}(\partial_{i}{\bf m}
\times\partial
_{j}{\bf m})\cdot{\bf m}$ can be expressed in the form
\begin{equation}
\delta\rho(r)=\frac {1}{4\pi r} \frac {d m^z}{d r},
\end{equation}
and the total charge is
\begin{equation}
Q=\int d^{2}r\, \delta\rho(r)=\frac {1}{2} \bigr[m^z(\infty)-m^z(0)\bigl].
\end{equation}
For a meron, the spin points up or down at the
core center and tilts away from the ${\bf\hat z}$ direction
as the distance from the core center increases.
Asymptotically it points purely radially in the ${\bf\hat x} -  {\bf\hat y}$
plane.
Thus the topological charge is $\pm \frac {1}{2}$ depending on the
polarity of core spin.
The variational function mentioned above corresponds to a vortex
with positive vorticity. In order to make a vortex with
negative vorticity (anti-vortex), we need to apply the space-inversion
operation to the vortex solution. Since topological charge is a pseudo-scalar
quantity, it is odd with respect to parity.
Hence the general result for the topological charge of the four meron
flavors may be summarized by the following formula:
\begin{equation}
Q=\frac {1}{2} \bigr[m^z(\infty)-m^z(0)\bigl]\, n_{\rm v},
\end{equation}
where $n_{\rm v}$ is the vortex winding number.

Finite energy excitations
can be formed by pairs of merons with opposite vorticity.
It seems likely that under appropriate circumstances
the lowest energy charged excitations of the system
will consist of a bound pair of
merons.  (A skyrmion can be viewed as a closely bound pair of merons with the
same charge and opposite vorticity
and a meron can be viewed as half a skyrmion.)
The energy of a pair of merons with opposite vorticity
(but like charge)
separated by a distance $R$ is given by
\begin{equation}
E_{mp} = 2 E_{\rm mc} + \frac{e^2}{4R} + 2 \pi \rho_E \ln (R/R_{\rm mc})
\label{eq:mp}
\end{equation}
where $E_{\rm mc}$ is the core energy of an isolated meron
$R_{\rm mc}$ is the core radius of an isolated meron, and the
expression should be applicable only when $ R \gg R_{\rm mc}$.
Minimizing  this expression with respect to $R$ gives a
meron separation $R^* = e^2 / 8 \pi \rho_E$.  Using the expression
for the $\rho_E$ derived in the last section this gives $R^* \approx 6
\ell$ for $d/\ell \sim 1$.
Quantum corrections are expected to reduce $\rho_E$ so
this expression should give a lower bound on the meron separation.
This very attractive picture of the lowest energy charge carriers in the system
[see Fig.(\ref{fig:mpair})] will only apply when the meron separation
is larger than the meron core size (which is expected to be $\sim \ell$),
and its energy is lower than the energy of a conventional
quasiparticle excitation where a charge is added with pseudospin
directed in opposition to the local pseudospin order.  It is
clear that we should expect the meron core size to increase
as $d / \ell$ approaches zero and the mass term in the
energy expression becomes small since there is only a small
energy cost for pseudospins to point out of the $xy$
plane.  Hence the picture is not likely to apply for very small
$d/ \ell$.  Further work which estimates meron core energies
and radii will be necessary to substantiate this picture and
is currently in progress.
We note in passing that the above description
of charged vortex antivortex pairs can also be used {\it mutatis mutandis\/}
for neutral vortex antivortex pairs.  These will have a conserved
momentum and the neutral collective mode (discussed in Section
\ref{sec:colmodes}) will cross
over from spin waves to such `quasiexcitons' at large wave vectors
in analogy to what occurs in the single layer case.\cite{kallin}

As in the case of skyrmions, we can write down explicit
microscopic variational
wave functions for vortices (merons). We start with the simplest example:
a meron with vorticity +1 and charge $-{1\over 2}$ that has the smallest
possible core size:
\begin{equation}
\vert\Psi_{+1,-{1\over 2}}\rangle
=\prod_{m=0}^M({1\over \sqrt{2}}c^{\dagger}_{m\uparrow}
+{1\over \sqrt{2}}c^{\dagger}_{m+1\downarrow})\vert 0\rangle.
\end{equation}
Here $\vert 0\rangle$ is the fermion vacuum, $c^{\dagger}_{m\uparrow,
\downarrow}$
creates an electron in the upper (lower) layer in the angular momentum $m$
state in the LLL and $M$ is the angular momentum quantum number
corresponding to the edge.
The vorticity is +1 because far away the spin wave function is essentially
\begin{equation}
\chi(\phi)={1\over \sqrt{2}}\left(\begin{array}{c}
e^{-i\phi}\\
1\end{array}\right),
\end{equation}
where $\phi$ is the polar angle.
The charge is $-{1\over 2}$ because we have created a hole in the center of the
lower layer ($m=0\downarrow$ is unoccupied).
Since the spin is pointing up at the center, this agrees with
the spin-charge relation derived earlier. From the spin-charge relation we
know we can change the sign of the charge of a meron by changing the direction
of spins in the core region without changing the vorticity. This can be seen
explicitly from the wave function:
\begin{equation}
\vert\Psi_{+1,+{1\over 2}}\rangle
=c^{\dagger}_{0\downarrow}\prod_{m=0}^M
({1\over \sqrt{2}}c^{\dagger}_{m\uparrow}
+{1\over \sqrt{2}}c^{\dagger}_{m+1\downarrow})\vert 0\rangle.
\end{equation}
This state has charge $+{1\over 2}$ because we have put in an electron in the
$m=0$ state in the lower layer. Obviously what we did is to flip the spins in
the core region to the down direction without changing the vorticity of the
meron, in the spin texture language.

A general wave function that describes a meron with vorticity $k$ has the form

\begin{equation}
\vert\Psi_{+k,-{k\over 2}}\rangle
=\prod_{m=0}^M
(a_m c^{\dagger}_{m\uparrow}
+b_m c^{\dagger}_{m+k\downarrow})\vert 0\rangle.
\label{eq:yangvor}
\end{equation}
This meron has charge $-{k\over 2}$ and we have assumed $k>0$. Generalization
to other cases is trivial. $a_m$ and $b_m$ are parameters
that satisfy
\begin{eqnarray}
|a_m|^2+|b_m|^2=1\nonumber\\
\lim_{m\rightarrow\infty}{a_m \over b_m}=e^{i\phi_0}
\end{eqnarray}
where $\phi_0$ is a constant. By adjusting the asymptotic behavior of
$a_m$ and $b_m$ (in particular their ratio,
 Eq.(\ref{eq:yangvor}) can also describe skyrmions and other
charged objects, while adjusting these coefficients
 in the core region one can modify the spin configuration there.
Since (\ref{eq:yangvor}) is a single Slater determinant, we can
calculate its energy using the original Hamiltonian directly without using the
effective
energy functional derived earlier. After subtracting the energy of the
fully spin-polarized state, we can show that the vortex energy diverges
logarithmically with the system size as would be expected for an XY system:
\begin{equation}
\Delta E_k\sim\gamma k^2\ln M
\end{equation}
where
\begin{equation}
\gamma=\sum_{l=0}^{\infty}{(-1)^l{2l+1\over 8}v^{\rm E}_l}.
\end{equation}
Here the $v^{\rm E}_l$ are Haldane's pseudopotential parameters
for the interlayer interaction\cite{pseudo}, and we have
neglected terms that are finite. If we use the energy functional, we find the
divergent part of the vortex energy is
\begin{equation}
\Delta E_k\sim{\pi\over 2}\rho_{\rm E} k^2\ln M
\end{equation}
implying
\begin{equation}
\rho_{\rm E}=\sum_{l=0}^{\infty}{(-1)^l{2l+1\over 4\pi}v^{\rm E}_l}.
\label{spinpseud}
\end{equation}
One can easily show that this expression of $\rho_E$ agrees with previous
expression exactly. The present expression in terms of pseudopotential
parameters may be more useful in finite size numerical calculations.
\cite{stability}

\section{Collective modes and Pseudospin Response Functions}
\label{sec:colmodes}

In this section we combine the equation of
motion for the spin-textures (Eq.[\ref{eq:eqmotion}]) and the
spin-texture energy functional (Eq.[\ref{eq:totaleng}])
to calculate the pseudospin linear response\cite{renndbl,jasonhononlinear}
 functions for $d/ \ell \ne 0$.
We take the pseudospin of the system to be polarized in the
${\bf \hat x}$ direction and consider the linear response to a time--
and space--dependent Zeeman field in the ${\bf\hat y} - {\bf\hat z}$ plane.
Fourier transforming with respect to both time and space we find that
\begin{equation}
\left(\begin{array}{cc}
- i \omega & - { 4 \pi  \over \nu } ( 2 \beta + q^2 \rho_A) \\
{ 4 \pi \over \nu } ( q^2 \rho_E) & - i \omega \\
\end{array} \right)\;
\left(\begin{array}{c} m_y \\ m_z \\ \end{array} \right)
=
\left(\begin{array}{c} - h_z \\ h_y \\ \end{array} \right)
\label{eq:resfunc}
\end{equation}
where $h_y$ and $h_z$ are the Fourier coefficients of the
pseudospin magnetic field at frequency $\omega$ and wavevector
$q$.  Physically $h_z$ corresponds to a time-- and space--dependent
bias potential between the two wells, while
$h_y$ could arise from a space-- and time--dependent interlayer
tunneling amplitude.  We see immediately that the response is singular when
\begin{equation}
\omega^2 = \omega_{\rm cm}^2 \equiv \left({ 4 \pi \over \nu }\right)^2
\left[2 \beta + q^2 \rho_A\right]  q^2 \rho_E.
\label{eq:colmode}
\end{equation}
Here $\omega_{\rm cm}$ is the frequency of a
long-wavelength collective mode of the system.  For the
$d/ \ell = 0$ case $,  \beta = 0 $, $\rho_A = \rho_E =\rho_s^0$,
and the collective mode frequency reduces to the result
obtained previously for the spin-wave collective mode of isotropic
ferromagnets. [$\omega_{\rm cm}  = 4 \pi q^2 \rho_s^0 / \nu $.]
The collective mode corresponds to a spin-precession whose
ellipticity increases as the long-wavelength limit is
approached.  The presence of the mass term ($\beta \ne 0$)
changes the collective
mode dispersion at long wavelengths from quadratic to
linear.  In the limit of small $q$
\begin{equation}
\omega_q={ 4\pi \over \nu} \sqrt{2\beta\, \rho_E }\,\, q.
\end{equation}

We can solve Eq.[\ref{eq:resfunc}] for the frequency and
wave vector dependent linear response to bias potential between the
two wells.  The result is
\begin{equation}
\chi_{zz} (q, \omega) = { 4 \pi q^2 \rho_E / \nu
\over \omega^2 - \omega_{\rm cm}^2 }.
\label{eq:chizz}
\end{equation}
It is interesting to compare this with formally exact
relations for this response function which can be obtained
from the microscopic Hamiltonian of double layer systems.
We first note that $\chi_{zz}$ is related to the dynamic structure factor,
\begin{equation}
s_{zz} (q, \omega) \equiv \sum_n | \langle \Psi_n  | S^z_q | \Psi_0
\rangle |^2  \delta \left( \hbar \omega - (E_n -E_0)\right),
\label{eq:szzform}
\end{equation}
by
\begin{equation}
s_{zz} (q, \omega) = -{ 1 \over \pi} {\rm Im} \chi_{zz} ( q, \omega + i \eta) .
\end{equation}
Here $| \Psi_n \rangle$ is an exact eigenstate of the double-layer
system.
Our result for $\chi_{zz}(q, \omega) $ thus implies that at
long wavelengths
\begin{equation}
s_{zz} (q, \omega) =
{ q  \over 2} \sqrt{ \rho_E \over  ( 2 \beta + q^2 \rho_A) }
\delta\left( \omega - \omega_{\rm cm} (q)\right) .
\label{eq:szzexpl}
\end{equation}
Some frequency moments of this structure factor can be related to
ground state correlation functions of the double-layer system.
The first moment gives the oscillator strength.  From Eq.[\ref{eq:szzexpl}]
we find that
\begin{equation}
f_{zz} (q, \omega) \equiv \int_0^{\infty} d \omega \omega s_{zz} (q, \omega)
 =  { 2 \pi \over \nu } q^2 \rho_E.
\label{eq:fsumrule}
\end{equation}
This result agrees with results for this moment calculated
previously\cite{renn,fsumrule} directly from the microscopic Hamiltonian as
we can confirm using Eq.[\ref{eq:estiffness}].
We see that the oscillator strength vanishes like $q^2$ as expected for an
ordinary superfluid, however the coefficient is proportional to $\rho_E$
and hence is non-universal. [The remaining oscillator strength is found
in a high frequency
collective mode lying below $omega_c$ by an amount proportional to $\rho_E$.
as shown below in Eq.(\ref{proj-sma}).]

The zeroth moment of the dynamic structure factor
gives the static structure factor:
\begin{equation}
s_{zz} (q) \equiv \int_0^{\infty} d \omega s_{zz} (q, \omega)
= \langle \Psi_0 | S^z_{-q} S^z_q | \Psi_0 \rangle ={ q  \over 2} \sqrt{
\rho_E \over  ( 2 \beta + q^2 \rho_A) } .
\label{eq:staticszz}
\end{equation}
We note that the static structure factor vanishes linearly with $q$
as $q$ goes to zero.  This property illustrates a qualitative
difference\cite{lopez} between the ground state of the double-layer system
at $d/ \ell \ne 0$ and the ground state
in the $d / \ell =0$ limit for which $s_{zz} (q, \omega)$ approaches
a constant as $q$ goes to zero.  This property is analogous to what happens
in a repulsively interacting Bose gas such as $^4$He in which the
structure factor vanishes linearly with $q$.  According to Feynman's
single-mode approximation picture, this is required in order to achieve
a linearly dispersing Goldstone mode.\cite{feynman}  The proper structure
factor can be included to improve the variational ground state wave function
by means of the Jastrow ansatz
\begin{equation}
\Psi =
\exp\left\{-\sum_q \frac{\lambda}{q} {\bar S}^z_{-q}{\bar S}^z_q\right\}
\prod_j \left(\begin{array}{c} e^{i\varphi/2} \\ e^{-i\varphi/2}
\\ \end{array} \right)_j \Psi_{V},
\end{equation}
where $\lambda$ is a variational parameter.
The incompressible $mmn$
states with $n\ne m$ discussed in section \ref{sec:chern-simons},
have the property\cite{fsumrule} that $s_{zz}(q) \sim q^2$ at
long wavelengths.  This property holds for the ground state whenever there
is an excitation gap and additional Jastrow factors are
not required to capture the correct long length scale fluctuations.

For $d/\ell \ne 0$ the mass term suppresses long length
scale fluctuations in $S^z$ which
measures the difference of the density in the two-layers.
The minus-one moment of $s_{zz} (q, \omega)$ is proportional to
the static response function:
\begin{equation}
\chi_{zz}(q,\omega=0) = { \nu \over 4 \pi (2 \beta + \rho_A q^2 ) }
\label{eq:chistatic}
\end{equation}
Note that $\chi_{zz} (q,\omega=0) $ diverges  in the $d / \ell$ limit
because of the broken $SU(2)$ symmetry in the ground state of
the $d/ \ell = 0$ system.

The collective mode dispersion can also be obtained from a
Lagrangian formulation which may be useful in describing
the thermodynamics of the system.  The partition function can be expressed by
\begin{equation}
Z=\int{D{\bf m}\, e^{-S^{E}[{\bf m}]}}
\end{equation}
where the Euclidean action is
\begin{equation}
S^{E}[{\bf m}]=
\int_{0}^{\beta}{d\tau}\biggl\{\int{d^{2}r[-i\frac{\nu}{4\pi}{\bf A}
({\bf m})\cdot\partial_{\tau}{\bf m}]}+E[{\bf m}]\biggr\}.
\end{equation}
The massive $m^z$ field is coupled to the massless field through the time
derivative term and the constraint $|{\bf m}|=1$.
For simplicity, we assume that the pseudospins are aligned along the
${\bf\hat{x}}$
direction {\it or\/}
 we concentrate on a local patch of the spin texture,
where the pseudospins are almost aligned along the ${\bf \hat{x}}$ direction.
Using the constraint,
we can express $m^{x}$ in terms of $m^{y}$ and $m^z$. We only
keep terms quadratic in $m^y$ and $m^z$.
The monopole vector potential is given in
a convenient gauge by ${\bf A}\simeq(0,-m^{z},m^{y})/2$.
Previously we took the long wavelength limit to obtain the local
action in real space since we were interested in results which would
become exact in the limit of extremely smooth spin
textures.  Here we generalize our discussion in order to obtain
an approximation for the full spectrum of
collective modes.  The calculations are identical to those detailed
in Section \ref{sec:action} except that we do not take the
long wavelength limit.  Here we report results only
for the case of experimental interest, $\nu=1$, where
the expressions take a somewhat simpler form.  For small $m_z$ and $m_y$
it is straightforward to integrate out the massive $m^{z}$ field
since the different momentum components decouple at the gaussian level
\begin{eqnarray}
e^{-S_{eff}^E[m^y]} &=& \int Dm^z \exp\biggl\{-\sum_{\omega_n,q}  m^{z}
(-\omega_n,-q)\omega_n m^y(\omega_n,q)/ 4 \pi \nonumber\\
&& + D_z(q)\vert m^z(\omega_n,q)\vert^2 +D_y(q) \vert m^y
(\omega_n,q)\vert^2\biggr\}\; ,
\end{eqnarray}
yielding
\begin{equation}
S_{eff}^{E}[m^{y}]=\sum_{\omega_n,q}{\biggl(\; {\omega_n^2\over 64
\pi^2 D_z(q)}
+D_y(q)\biggr) \vert m^{y}(\omega_n,q)\vert^2} .
\end{equation}
\noindent
We can read off the collective mode frequency from this expression:
\begin{equation}
\omega_k=8 \pi \sqrt{D_z(k)D_y(k)},
\label{chad2}
\end{equation}
where,
\begin{eqnarray}
D_z(k) &=& \frac {1}{8 \pi } \biggl\{\frac {1}{\pi \ell^2} V_z(k)
e^{-\frac {k^2}{2}}\nonumber\\
&+&  \int \frac {d^2 q}{(2\pi)^2} V_E(q) \exp ( -|q|^2/2)
-\int \frac {d^2 q}{(2\pi)^2} V_A(q)
\exp( - |q|^2/2)  e^{i k \wedge q}\biggr\} ,
\end{eqnarray}
and
\begin{equation}
D_y(k) =\frac {1}{8 \pi } \int \frac {d^2 k}{(2\pi)^2} V_E(q) \exp(
-|q|^2/2 ) ( 1
- e^{i k \wedge q})\;.
\end{equation}
\noindent
This spectrum agrees exactly with the dispersion relation
in Refs.[\onlinecite{gapless,ahmz1}],
where the excitation energy for the single magnon state was
obtained using the time-dependent Hartree-Fock approximation
and using a single mode approximation combined with the Bogoliubov
transformation.   In the long wavelength limit the dispersion
relation reduces to the result discussed above.

\section{Kosterlitz-Thouless Phase Transition and Spin-Channel
Superfluidity}
\label{sec:kostthou}

The linearly dispersing gapless mode discussed in the previous section
and the absence of gapless charged excitations suggest that
the system should show superfluid behavior
in the pseudospin channel as has been noted previously\cite{ezawa,wenandzee}.
To make this suggestion more concrete we evaluate the linear
response of the system to opposing electric fields in
the two layers.  Combining Eq.[\ref{eq:chizz}] with the continuity
equation for the ${\bf\hat z}$ component of pseudospin we find that
\begin{equation}
\sigma_{zz}(q, \omega) = { e^2 \omega \chi_{zz} (q, \omega) \over i q^2}
= { 4 \pi e^2 \rho_E \omega / i \hbar^2 \nu
\over \omega^2 - \omega_{\rm cm}^2}.
\label{eq:sigzz}
\end{equation}
This conductivity gives the difference between
oppositely directed charge currents which flow in the two layers in
response to oppositely directed electric fields.  (Note
that no net current will be induced by such electric fields).
In the long wavelength limit this leads to a frequency dependent
conductivity equivalent to that of a superfluid with
superfluid density proportional to $\rho_E$:
\begin{equation}
\sigma_{zz}(\omega) = { 4 \pi e^2 \rho_E \over i \nu \hbar^2 \omega}.
\label{eq:sigzzom}
\end{equation}
Note that by the Kronig-Kramers relation this conductivity must
have a real part which is proportional to a $\delta$ function at
zero frequency.   We remark that the superfluid property requires
not only a gap for charged excitations and the linearly
dispersing collective mode but also a total oscillator strength
which vanishes as $q^2$ at long wavelengths.  For the conductivity associated
with response to electric fields in the same direction in the two layers, the
total oscillator strength associated with intra-Landau-level
excitations vanishes\cite{fsumrule} as $q^4$ and the the collective mode
has a gap at long wavelengths.  These properties lead to a
$\sigma( q, \omega)$ which vanishes as $q^2$ in the long wavelength
limit leading to the quantum Hall effect rather than to
superfluidity.

The above analysis is dependent on our analysis of the response
functions of the double-layer system which does not include
thermal fluctuations.  At finite temperatures both meron
and pseudospin collective mode thermal fluctuations have to
be accounted for.  As in other two-dimensional superfluids
the linear response conductivity is still expected to vanish
below a finite temperature.
As we have discussed previously the low-energy excitations of
the double-layer system consist of highly elliptical precessions
of the spin about the direction of the local order parameter.
It follows from a continuity equation and the equation of motion
for the spin textures that the conserved number current corresponding to the
$\hat z$ component of the pseudospin is related to the pseudospin by
the usual minimal coupling prescription
(see Appendix  for further discussion of this result)
\begin{equation}
J_{zz} = { 2 \rho_E \over \hbar } {\bf \nabla \phi}
\label{eq:spincurrent}
\end{equation}
where the factor of two arises from the definition of $J_{zz}$
 as the difference of the number currents in the two layers, and
$\phi$ is the projection of the pseudospin orientation onto
the $\hat x - \hat y$ plane.  $\phi$ plays the same role as the
phase of the superconducting order parameter in a two-dimensional
superfluid and pseudo-spin channel superfluidity will be
coincident with pseudospin easy-plane ferromagnetism.  Likewise, as we
discuss further below and in Appendix , the divergent superfluid
conductivity implies zero pseudospin Hall resistivity below the KT
temperature.

In this system the Kosterlitz-Thouless (KT) phase
transition\cite{kt} which separates the
superfluid and normal states is expected to
be\cite{sfcaveat} associated with the unbinding
of meron pairs of opposite charge and opposite vorticity.
In order to analyze the KT transition, it is better to work in real space,
since the KT transition is controlled by the large-distance physics and
the short-distance details can be effectively taken into account by
the vortex core energy.  In the limit of a large mass term
($ \beta \ell^2 \gg \rho_E$) fluctuations in $m_z$ will be small.
After integrating out the massive field and finite frequency
fluctuations in a gaussian approximation, we obtain an effective XY model
\begin{equation}
S_{eff}^E = {\beta \over 2}\rho_{E}\, \int{d^{2}r}\, (\nabla\varphi )^2,
\end{equation}
where $\varphi$ is an angle denoting the direction of the spin in the
XY plane.
We know that this model undergoes a Kosterlitz-Thouless phase
transition associated with the unbinding
of bound vortex pairs\cite{kt} at approximately
\begin{equation}
T_c = {\pi\over 2} \rho_{E}.
\end{equation}
For a 2D XY model corrections to this expression for
$T_c$ arise from finite temperature spin-wave and
vortex-antivortex polarization renormalizations of
the effective spin stiffness at long distances.  The magnitude
of the corrections depends on details of the short-distance
physics.  For the 2D nearest-neighbor-coupling XY model
on a square lattice\cite{mc}
whose short-distance physics we believe to be similar to that of double-layer
systems
\begin{equation}
T_c \sim 0.90 \rho_{E}.
\label{eq:tcxy}
\end{equation}
(We note however that the double-layer system possesses charged
excitations whose energies are larger but of the same order
as the meron core energies.)
In the present case $T_c$ should be further reduced, especially
as $d/ \ell$ goes to zero, because of fluctuations out of the
${\bf\hat x} - {\bf\hat y}$ plane.
However, numerical studies of the anisotropic
$O(3)$ model on a square lattice\cite{klomfass} demonstrate that these
corrections are not very important except in the limit of extremely weak
anisotropy.  For example $T_c$ is reduced by a less than a
factor of two compared to Eq.[\ref{eq:tcxy}] even for parameters
which correspond to $\rho_E \ell^2 / \beta  \sim 30$.  Comparing with
Fig.(\ref{fig:massstiffnu1}) we see that such weak anisotropies
occur in double-layer systems only for $d / \ell < 0.3$, a regime
which is not experimentally accessible.  (Quantum fluctuations
only increase the anisotropy by decreasing $\rho_E$.) In the
following section we provide a quantitative estimate of the
temperature scale expected for $T_c$ by combining Eq.[\ref{eq:tcxy}]
with estimates of $\rho_E$ which include the effect of quantum
fluctuations.  As we discuss there, the principle source of
uncertainty in the $T_c$ estimate comes from attempting to
estimate $\rho_E$.

It should be noted that the U(1) symmetry responsible for the XY model
physics and the KT transition is robust under application of a bias voltage
which induces a charge imbalance between the layers.  Within the
pseudospin picture, the spins simply tilt slightly out of the XY plane
in the positive or negative $z$ direction.  This will reduce the component
of the spin in the XY plane and hence may lower the KT temperature slightly
but it will not (for weak imbalance) destroy the ordering transition.  This
is in sharp contrast to the behavior for $mmn$ states with $m>n$ which
require precise charge balance for their existence.  This point is
discussed in more detail in Section \ref{sec:exactdiag} below.

In an ordinary superconducting film, the linear response conductivity
is infinite below the KT transition temperature.
More precisely the voltage-current relationship obeys:
\begin{equation}
V \sim I^{p(T)}
\end{equation}
where the exponent $p(T) = 1$ for $T>T_{\rm KT}$ and $p(T)>3$ for
$T<T_{\rm KT}$.  The exponent has a universal jump and changes
discontinuously from 1 to 3 at the transition.
The critical current is zero, {\it i.e.}
there is always a finite voltage for any finite
current.  At any finite temperature
Bogoliubov quasi-particles are present, however these do not
produce dissipation because the current is carried by
the superfluid and the voltage is (essentially) zero.  The superfluid
`shorts out' the normal fluid and so there is no electric field to produce
motion of the normal fluid.  The present problem is richer because
of the existence of two layers in which the current can flow and
because of the possible presence of a Hall voltage in the presence
of currents.  It is instructive to define phenomenological\cite{renndbl}
transport coefficients (for the case of two identical layers) as follows:
\begin{equation}
\left(\begin{array}{c} E_1^x \\ E_1^y \\ E_2^x \\ E_2^y \\ \end{array}
\right) =
\left(\begin{array}{cccc} \rho_{xx}^A & \rho_{xy}^A & \rho_{xx}^E &
\rho_{xy}^E \\
-\rho_{xy}^A & \rho_{xx}^A & -\rho_{xy}^E & \rho_{xx}^E \\
\rho_{xx}^E & \rho_{xy}^E & \rho_{xx}^A & \rho_{xy}^A \\
-\rho_{xy}^E & \rho_{xx}^E & -\rho_{xy}^A & \rho_{xx}^A \\
\end{array}\right)
\left(\begin{array}{c} j_1^x \\ j_1^y \\ j_2^x \\ j_2^y \\
\end{array}\right).
\label{eq:transport}
\end{equation}
Here the numerical subscripts label the two layers of the double-layer
system. (Note that this phenomenology reflects the fact that the
transport coefficients for the sum and differences of the currents
or electric fields decouple.)

In general there are four independent
transport coefficients, allowing for contributions to both
dissipative and Hall electric fields due to currents flowing in the
same layer, $ \rho_{xx}^A$ and $\rho_{xy}^A$, and the interlayer
or drag\cite{drag} resistivities coming from currents flowing in the opposite
layer, $ \rho_{xx}^E$ and $\rho_{xy}^E$.  The ratios of the
electric field sums and
differences to the current sums and differences are given
by $\rho^A +  \rho^E$ and $\rho^A - \rho^E$ respectively for both
dissipative and Hall fields.  The matrix of conductivity coefficients
may be related to the matrix of resistivity coefficients by
inverting Eq.[\ref{eq:transport}].  We have argued above that the
dissipative dc conductivity coefficient for current differences is infinite
when $T < T_{KT}$.
This implies that both the Hall and dissipative resistivities
are zero for $ T < T_{KT}$, {\it i.e} that
$\rho_{xx}^E = \rho_{xx}^A$ and $\rho_{xy}^E = \rho_{xy}^A$.
Below the Kosterlitz-Thouless temperature there are only two independent
dc linear transport coefficients.  Moreover we know a great deal
about these two transport coefficients because of the
quantum Hall effect.  When identical electric fields exist in
each layer the total Hall conductance is nearly exactly quantized at
low temperatures ($ \sigma_{xy}^A + \sigma_{xy}^E \approx \nu e^2 / h $)
and the dissipative conductance is activated ($ \sigma_{xx}^A +
\sigma_{xx}^E \sim \sigma_0 \exp ( - \Delta / 2 k_B T)$).  Here
$\sigma_0$ is a non-universal constant and $\Delta$ is the gap for
making charged particle-hole pairs; these charged objects
are probably the meron pairs discussed previously.
In terms of the linear resistivity
matrix of Eq.[\ref{eq:transport}] we conclude that for low temperatures
\begin{equation}
\rho_{xx}^A = \rho_{xx}^E = { \sigma_0 \exp (- \Delta / 2 k_B T )
\over 2 \nu^2 (e^2 / h)^2 }
\label{eq:rhoxx}
\end{equation}
and
\begin{equation}
\rho_{xy}^A= \rho_{xy}^E = { - h \over 2 \nu  e^2 }.
\end{equation}
Note that this implies the occurrence of remarkable cross-talk
phenomena.\cite{renndbl,duan}  For example if current is injected in the
${\bf \hat x}$
direction in one layer but no current flows in the other layer,
a quantized Hall field whose value corresponds to the density
{\it per layer} will appear in both layers.

Ho\cite{jasonho} has recently considered the question
of the stability of current flow in the spin channel.  He finds that the
existence of oppositely directed electric fields in each layer induces
phase twists which can be relieved only by steady nucleation of topological
defects in close analogy with textures induced in superfluid $^3$He.  We
agree that such effects will occur at finite current densities, however
these results disagree with our findings above that the linear response
$\sigma_{xx}$ is infinite and $\rho_{xy}$ is zero.  We believe that this
discrepancy has two origins:  the finite easy-plane anisotropy is essential
to the analysis of this problem and the fact that there are subtleties
associated with the question of the existence of uniform current carrying
states in the lowest Landau level.  [This latter point is discussed in
detail in the Appendix.] Without easy plane anisotropy, the SU(2) symmetry
prevents the XY order necessary for superfluidity.  However since even
{\it with\/}
easy plane anisotropy, the XY order is only algebraic, the critical current
density is zero as discussed above.  For any finite current density there will
indeed be dissipation due to generation of topological defects.

\section{Exact Diagonalization Studies}
\label{sec:exactdiag}

In this section we discuss some microscopically based investigations
of the properties of double-layer systems at $\nu = 1$.
We start by discussing some studies using the finite-size
exact-diagonalization in the spherical geometry\cite{fdmhz1}.
As we have emphasized previously at $d / \ell = 0$,
where intra-- and inter--layer Coulomb potentials are the same,
the electron-electron interaction term in the Hamiltonian
is invariant under rotations in pseudospin space.
Eigenstates of the Hamiltonian $H$ can be simultaneous
eigenstates of any one component of the total pseudo-spin
operator.  States with pseudo-spin quantum number $S$ will
have degeneracy $2S+1$.  As discussed in Section \ref{sec:analogy}
the ground state at $\nu=1$
is a pseudo-spin eigenstate with $S=N/2$ where $N$ is the
number of electrons.)
At finite layer separations, the inter-layer interactions will
be weaker than intra-layer interactions and the broken
symmetry in the ground state is reduced from SU(2) to
U(1).  Microscopically this corresponds to the fact that
in the ground state, electrons within the
same layer will be more strongly correlated than electrons
in different layers in order to minimize the total
Coulomb energy.  For sufficiently widely separated
layers we will provide evidence that the ground state
no longer has broken symmetry.
This problem has been studied before through examination of the ground state
wavefunction,\cite{gsnumamd,DYAHM} and a ground state level crossing
was found to occur in the vicinity of $d/\ell=1.5$.
In this section we will discuss another
attempt to estimate the layer separation at which this quantum
phase transition takes place.  In addition we will discuss
rough estimates obtained for the charge-gap and for the
parameters of the spin-texture energy functional.

The model we consider consists of two
two-dimensional electron systems separated by a distance $d$.
For the sake of definiteness the spread of the electron wave function
in the perpendicular direction in each layer is neglected.
(Such effects are easily accounted for if the geometry of
a particular sample is known; we assume that it will normally
be possible to define an effective layer separation for each sample.)
We start by considering the case of greatest interest where
each layer has the same number of electrons and define the
zero of energy by placing neutralizing non-responding background charge
backgrounds on each layer.  We parameterize the intra-- and inter--layer
interactions in terms of the Haldane
pseudopotential parameters\cite{fdmhz1} for this model.
Calculations were performed for
$N=4,6,8,10$ at a variety of layer separations.

We first discuss finite size estimates of the chemical potential
dependence at $\nu = 1$.  For a finite system of electrons on the
surface of a sphere the chemical potential jump is expected
to occur when $N=N_{orb}=N_{\phi}+1$. Here $N_{orb}$ is the number
of orbitals per Landau level when $N_{\phi}$ flux quanta pass
through the surface of the sphere.  The
chemical potential discontinuity is given by the limit as
$N \to \infty$ of
\begin{equation}
\Delta\mu=E(N_\phi,N+1)+E(N_\phi,N-1)-2E(N_\phi,N),
\label{equ:dmu}
\end{equation}
where $N_\phi=N -1$. In Fig.(\ref{fig:fz1}), the finite
size estimate of $\Delta\mu$ for the $\nu=1$ state is
shown as a function of layer separation $d$
for $N=8$ and $N=10$.  $\ \Delta\mu$ decreases continuously
as $d$ increases.  The finite size corrections can be crudely
inferred from comparisons of results for
$N=8$ and $N=10$.
For a system size increase from $N=8$ to $N=10$,
$\ \Delta\mu$ increases slightly for $d <1.5 \ell$, and
decreases substantially for $d>1.5 \ell$.  In all likelihood this
result indicates that in the thermodynamic limit ($N=\infty$),
there is no chemical potential discontinuity for $d > \sim 1.5 \ell$.
For layer separations larger than this, it seems likely
that the properties of the
double layer system should be similar to those of two isolated
layers with Landau level filling factor $\nu =1/2$,
which are of great interest in their own right.\cite{nuhalf}
The existence of an intermediate state, best viewed as a
quantum disordered easy-plane ferromagnet, is an interesting
speculative possibility.

We have argued that the charged objects, at least
away from $d / \ell =0$ consist of meron pairs with opposite vorticity.
Since we expect the meron core energy and the pseudospin stiffness
to both vanish when the order parameter vanishes, we expect that
the chemical potential discontinuity should vanish at the
critical layer separation where the zero-temperature
phase transition to a disordered state occurs.
Thus these results also provide an estimate of the
layer separation at which the quantum phase-transition takes
place.   We can get an estimate of the finite-size errors
in these results at small $d/ \ell= 0 $ where the value
of the chemical potential discontinuity is known\cite{sondhi}
exactly for $\nu=1$.
At $\nu =1 $ and $d/ \ell =0$, the ground state many-body
wavefunction {\it is} a single Slater determinant so that there
are no quantum fluctuation corrections to the Hartree-Fock
value for the spin-stiffness given in Section \ref{sec:action}.
The chemical potential discontinuity is therefore given exactly for
$ N = \infty$ by $ \Delta \mu = (e^2 / \ell) (\pi / 8)^{1/2}$.
This value is marked by an asterisk in Fig.(\ref{fig:fz1})
and compares well with the finite-size estimates.

To get further insight into the system, we have attempted to
estimate the dependence of the ground-state
order parameter ($M$) and
the easy-plane pseudospin magnetic susceptibility ($\chi$)
on $d / \ell$ for $\nu =1$.  We normalize the order parameter
so that it has the value $1$ at $d /\ell$ =0.
The order parameter is defined by
\begin{equation}
M \equiv \lim_{t \to 0} \big[ \lim_{N \to \infty}
-(1/N){d\over dt}E(N_\phi,N)| \big]
\label{eq:ordparm}
\end{equation}
and the magnetic susceptibility is defined by
\begin{equation}
\chi \equiv \lim_{t \to 0} \big[ \lim_{N \to \infty}
-(1/N){d^2\over dt^2}E(N_\phi,N)| \big]
\label{eq:suscep}
\end{equation}
where $t$ is the inter-layer tunneling amplitude (which acts like a
Zeeman field on the pseudospins).
Some typical\cite{more} results for the dependence of the ground
state energy on tunneling amplitude are shown in
Fig.(\ref{fig:vsN}) and Fig.(\ref{fig:vsd}).
The solid lines in these figures are interpolations between
energies calculated at a series of equally spaced $t$ values
{\it not} including $t =0$.  In a finite size system it is
clear that the ground state
energy always will have a quadratic dependence on tunneling
amplitude until the total `Zeeman'
coupling becomes comparable to the smallest excitation energy
of the system so that the order of limits in Eq.[\ref{eq:ordparm}]
and Eq.[\ref{eq:suscep}] is essential.  The lowest energy
excitations of the system will be long wavelength collective
modes and we can expect to estimate the dependence of the
zero-temperature order parameter and susceptibility on $d / \ell$
from the dependence of the ground state energy on $t$ in the regime
where the $t$ is small compared to $e^2 / \ell$ but the
Zeeman energy is large compared to the minimum excitation energy
of the system.  Using our result for the collective mode energy
and the correspondence between linear momenta and
angular momenta on a sphere ($ k \sim L / R$ where $R$ is the
sphere radius) we estimate that the minimum value of $t$ we can
consider is
\begin{equation}
t_{\rm min} \sim { 4 \pi \sqrt{ \rho_E \beta } \over \nu\ell  N^{3/2} }.
\label{eq:tmin}
\end{equation}
For $\nu =1$, $\rho_E$ is strongly renormalized downward
by quantum fluctuations and we expect it to drop to zero
as $d / \ell$ approaches $\sim 1.5$ while $\beta$ starts
from zero at $d / \ell= 0$,  rises quadratically and is less
affected by quantum fluctuations.  Comparing with
Fig.(\ref{fig:massstiffnu1}) we estimate that for $d / \ell \sim 1$
and $N=10$, $t_{\rm min} \sim 10^{-3} ( e^2 / \ell) $; $t_{\rm min}$
will approach
zero for $d / \ell \to 0$.  These rough estimates are consistent
with the crossovers from quadratic to linear dependence on
$d$ seen in Fig.(\ref{fig:vsN}) and Fig.(\ref{fig:vsd}).
We have estimated the magnetization and susceptibility by
interpolating between calculated energy values and
evaluating the derivatives at $d \sim 2.5 \times 10^{-3}$.
The results are shown in Fig.(\ref{fig:mag}) and
Fig.(\ref{fig:chi}).  The magnetization decreases
continuously with $d$ and appears to vanish in the limit of large
$N$ at a critical layer separation consistent with the estimate
obtained by looking at the charge gap.  The finite-size susceptibility
estimate becomes large near where the order parameter vanishes
as expected\cite{attempt}.

The order parameter is an important parameter which will
appear in the spin-texture energy functional in the presence
of interlayer tunneling as discussed in Ref.[\onlinecite{II}].
We discuss here attempts to estimate parameters $\beta$ and
$\rho_E$.  Fig.(\ref{fig:beta}) shows the dependence on the layer
separation of the increase
in the ground state energy when one and two electrons are
transferred from one layer to another.
According to the spin-texture energy functional this energy should
be given, in the limit of large systems, by
\begin{equation}
\Delta E = \beta (d/ \ell) \ell^2 2 \pi\left[\frac{(\Delta N)^2}{N} \right].
\label{eq:betafs}
\end{equation}
The finite size exact diagonalization calculations were performed
for $d/ \ell =0.5, d / \ell =1.0$, and $d / \ell = 1.5$ and the values
of $\beta \ell^2 / (e^2 / \ell) $
inferred by comparing with Eq.[\ref{eq:betafs}] were
$ 5.1 \times 10^{-3}$, $1.7 \times 10^{-2}$ and
$ 4.0 \times 10^{-2}$ respectively.  These values compared to
$ 2.0 \times 10^{-2}$, $4.0 \times 10^{-2}$ and
$6.0 \times 10^{-2}$ respectively for the Hartree approximation and
$ 7.3 \times 10^{-3}$, $2.2 \times 10^{-2}$ and
$3.8 \times 10^{-2}$ for the Hartree-Fock approximation results
derived earlier.  The quadratic dependence derived in the
Hartree-Fock approximation is apparent in the exact diagonalization
results.  The exact diagonalization estimates for $\beta$ demonstrate
that quantum fluctuation corrections to this quantity are quite
small and that the Hartree-Fock results shown in
Fig.(\ref{fig:massstiffnu1}) are quite accurate.  An important aspect
associated with the broken symmetry in the ground state at
$\nu =1$ is the fact that the chemical potential discontinuity
is not strongly influenced by the transfer of charge from
one layer to the other.  In the presence of a
bias potential the ground state pseudospin is tilted out of
the $\hat x-\hat y$ plane but the system still has a broken
$U(1)$ symmetry.  This situation contrasts with the case of
the double-layer quantum Hall effect which occurs at total
Landau level filling factor $\nu_T = 1/2$ where the chemical
potential discontinuity occurs only near equal layer
densities.\cite{lzhengunpub}

In Fig.(\ref{fig:colmodea}) and Fig.(\ref{fig:colmodeb}) we show
the dependence of the lowest excitation energy on wavevector for
a finite system of electrons on a torus with periodic boundary
conditions\cite{haldanetorus} for $d/ \ell =0.5$ and
$d / \ell =1.0$ respectively.  At $d/ \ell =0.5$ clear
indications of the linearly dispersing collective mode are
evident.  The velocity of the linearly dispersing collective
mode is related to the parameters of the spin-texture
energy by
\begin{equation}
E(k) = {k \ell} { 4 \pi \over \nu} \sqrt{ 2 \beta \rho_E}.
\label{eq:velocity}
\end{equation}
Reading off the velocity from Fig.(\ref{fig:colmodea}) and using the
value of $\beta $ obtained above from the exact diagonalization
calculation we estimate that for $d / \ell =0.5 $,$ \; \rho_E
 \approx 1.5 \times 10^{-2} (e^2 / \ell) $ compared to the
 Hartree-Fock value $\rho_E = 1.22 \times 10^{-2}$.  For
 $d/ \ell =1.0$ the linear dispersion of the collective mode
 is less evident and it is more difficult to estimate
the velocity accurately.  We have chosen to estimate the velocity
from the energy at the smallest finite-size system wavevector as
indicated in Fig.(\ref{fig:colmodeb}) from which we obtain
$\rho_E \approx 3.3 \times 10^{-3} (e^2 /\ell)$ compared to the
Hartree-Fock value $\rho_E = 4.22 \times 10^{-3}$.  At this value
it is evident that quantum fluctuations are decreasing the
value of $\rho_E$ as the critical layer separation is approached.

To obtain a rough quantitative estimate of the layer separation
dependence of the Kosterlitz-Thouless transition temperature, we
calculate the leading order corrections to the
Hartree-Fock ground state energy as a function of parallel
magnetic field $B_{||}$.
The renormalized spin stiffness $\rho_E^R$ is related to the
dependence of the ground state energy on field by the following
equation\cite{II}
\begin{equation}
\rho_E^R = \frac {d^2}{dQ^2}  E_A(Q)|_{Q\rightarrow 0}
\end{equation}
where $E_A$ is ground state energy per area and $Q=dB_{||}/\ell^2 Bp$.
Given $\rho_E^R$ and the fact that the mass term is not
strongly influenced by quantum fluctuations, we can estimate the
effect of fluctuations in $m^z$ on the transition temperature
by using results from numerical simulations\cite{epnlsig} of the classical
$O(3)$ `easy-plane' non-linear sigma model,
We find\cite{detailslater} the dependence of
$T_{KT}$ on  $d/\ell$ which is illustrated in Fig.(~\ref{fig:tckt}).
As the layer separation approaches zero, $U(1)$ symmetry is enhanced to
$O(3)$, and fluctuations of $m^z$ eventually become sufficiently important
that $T_{KT}$ decreases and approaches zero.
On the other hand for layer separation close to the critical one,
phase coherence between the layers is destroyed only
by quantum fluctuations.  Our results for $T_{KT}$ are in
good agreement with the $d/\ell$ dependence
of the temperatures at which features in the dissipative resistance
of double-layer systems have been seen\cite{Shayegan} by Lay {\it et. al.}.
These features might be associated with
the vanishing charge gap expected to occur in parallel
with the KT transition.

\section{Chern-Simons-Landau-Ginzburg Theory}
\label{sec:chern-simons}

An useful alternative way of understanding the physics of the quantized Hall
effect is based on the concept of composite bosons\cite{gm,zhk,read}.
Much of the physics of the single layer quantum Hall effect can be
described in terms of the Chern-Simons-Landau-Ginzburg theory\cite{zhk,zhang}.
 One starts with the observation that the problem
at odd denominator filling fraction can be mapped exactly to a problem
of bosons with an odd number of statistical
flux tubes attached to them. At the mean field level, the statistical
flux cancels the external magnetic field and one obtains a boson
superfluid. 
Treating the fluctuations above this
mean field within the RPA approximation
restores the gap and    
one obtains an incompressible
bosonic liquid.   
Both Laughlin's wave function\cite{laughlin} and the long wavelength
algebraic off-diagonal-long-order correlation function\cite{gm} can be derived
explicitly from the CSLG theory\cite{zhang}.

The CSLG formulation of the single layer quantum Hall system has been
extended both to the case involving electron spin\cite{leekane,sondhi} and
to the double layer case\cite{renndbl,wenandzee,ezawa,birman}
by a number of authors.
The Chern-Simons theory of the double layer system based on the
fermionic representation has also been constructed\cite{lopez}.
In the case of
double layer systems, there is both a statistical gauge interaction of the
composite bosons within the layers and between the layers. The action
for this problem is given by
\begin{eqnarray}
{\cal L}_\phi & = & \phi\sp\dagger_{\uparrow}
 (i \partial_t + A^0 - a^0_{\uparrow}) \phi_{\uparrow}
  - \frac{1}{2m^*} \left|\left(\frac{1}{i} \; {\bf\nabla}
  + {\bf A} - m{\bf a}_{\uparrow} - n{\bf a}_{\downarrow}
   \right) \phi_\uparrow \right|^2 + \frac{1}{4\pi} \;
\varepsilon\sp{\mu\nu\rho} \;
a^\mu_{\uparrow} \partial_\nu \; a^\rho_{\uparrow}
\nonumber \\
& & \mbox{} - \frac{1}{2} \int d^2y \; \delta\rho_{\uparrow}(x) \;
V_A (x-y) \; \delta\rho_{\uparrow}(y) + (\uparrow \rightarrow \downarrow)
  - \int d^2y \; \delta\rho_{\uparrow}(x) \; V_E (x-y) \;
  \delta\rho_{\downarrow}(y),
\label{action}
\end{eqnarray}
where $(\uparrow \rightarrow \downarrow)$ indicates the first four
corresponding terms with up and down labels interchanged.
Here $m$ is an odd integer. In the absence
of tunneling, the particles in the two different layers are distinguishable,
the relative phase winding between them can be either $0$ or $\pi$, therefore,
the integer $n$ can be either even or odd. At the mean
field level, the equations of motion are given by
\begin{equation}
{\bf\nabla} \times \; {\bf a}_{\sigma}(x) = 2\pi \rho_{\sigma}(x) \ \ \ ; \ \ \
{\bf A} = m {\bf a}_{\uparrow} + n {\bf a}_{\downarrow}.
\label{mean-field}
\end{equation}
When the electron densities of both layers are equal,
we see that the the filling factor has to be $\nu=2/(m+n)$ for these equations
to be satisfied.

{}From this CSLG formulation Halperin's wave function\cite{halperinz1}
for the $(mmn)$ state
can be derived in a fashion similar to the single layer case\cite{zhang}.  One
can decompose
the complex boson field in terms of the amplitude and the phase part,
\begin{equation}
\phi_{\sigma}(x) =
\sqrt{\bar{\rho}+\delta\rho_{\sigma}} \; e\sp{i\theta_{\sigma}(x)}.
\label{decompose}
\end{equation}
The Chern-Simons gauge field induces a long-ranged logarithmic
density-density interaction,
giving rise to the following effective Hamiltonian at the quadratic level:
\begin{equation}
H = \frac{1}{2}\; \frac{\bar{\rho}}{m^*} \sum_q \left[ \left( \frac{2\pi}
  {q} \right)^2 \left( (m^2 + n^2) \pi_\sigma(q) \; \pi_\sigma(-q) +
   4mn  \pi_\uparrow(q) \; \pi_\downarrow(-q) \right) +
     q^2 \;  \theta_\sigma(q) \; \theta_\sigma(-q) \right],
\label{eff-ham}
\end{equation}
where $\pi^{\sigma}_q$ is proportional to the density
$\delta\rho_\sigma(x)$ and is the conjugate variable of the phase, i.e.
\begin{equation}
[\theta_\sigma(q) , \; \pi_{\sigma'}(q')] = -i \;
\delta_{\sigma\sigma'} \delta(q+q').
\label{conjugate}
\end{equation}
This Hamiltonian is a direct sum of harmonic oscillator Hamiltonians for each
wave vector $q$.
By forming the sum and the differences of these operators, one can easily
diagonalize the Hamiltonian and find the ground state wave function
in terms of its dependence on the generalized coordinates (densities)
$\pi^{\sigma}_q$:
\begin{equation}
\Psi_0 [\pi_q] = \; \exp \; \left\{\frac{1}{4} \sum_q \frac{2\pi (m+n)}{q^2}
\; \pi^+_q \; \pi^+_{-q} + \frac{2\pi (m-n)}{q^2}
  \; \pi^-_q \; \pi^-_{-q} \right\}.
\label{wavefunction}
\end{equation}
One can express the density operators in terms of
the ordinary first quantized coordinates of the particles
\begin{equation}
\pi^+_q = \frac{1}{\sqrt{2}}
\sum_i \; \left\{e^{iqx^\uparrow_i}
+ e^{iqx^\downarrow_i}  - 2 \bar{\rho}\right\} \ \ \ ;
\ \ \ \pi^-_q = \frac{1}{\sqrt{2}} \sum_i \;
\left\{e^{iqx^\uparrow_i} - e^{iqx^\downarrow_i}\right\},
\label{density}
\end{equation}
and obtain the first quantized wave function
\begin{equation}
\Psi_0(x^\uparrow_i, x^\downarrow_j ) = \prod_{i<j} |x^\uparrow_i -
x^\uparrow_j|^m \;
  |x^\downarrow_i - x^\downarrow_j|^m \; |x^\uparrow_i - x^\downarrow_j|^n
  e^{-\frac{1}{4} \sum_{\sigma,i}|x^\sigma_i|^2}.
\label{halperin}
\end{equation}
This is nothing but Halperin's wave function for the $(mmn)$
state expressed in terms of
the composite boson variables. One can easily perform a singular
gauge transformation
back to the original electron coordinates and find explicitly
Halperin's wave function
expressed in terms of the original electrons.

{}From the effective Hamiltonian (\ref{eff-ham}) one can also derive
all static correlation
functions in the long wavelength limit. One example is the analogue
of the ODLRO
correlation function first introduced by Girvin and MacDonald\cite{gm} for
the single layer case. From (\ref{eff-ham}) one easily obtains
\begin{equation}
 <e^{i\theta_+(x)-i\theta_+(y)}> = |x-y|^{-(m+n)} \ \ \ ; \ \ \
 <e^{i\theta_-(x)-i\theta_-(y)}> = |x-y|^{-(m-n)}.
\label{ODLRO}
\end{equation}
Similarly, one can obtain the static density correlation function
in the long wavelength limit (for $m\ne n$)  
\begin{equation}
 <\pi^+_q \; \pi^+_{-q} > = \frac{q^2}{2\pi (m+n)} \ \ \ ; \ \ \
 <\pi^-_q \; \pi^-_{-q} > = \frac{q^2}{2\pi (m-n)}.
\label{static}
\end{equation}

We therefore conclude that the CSLG theory correctly gives the static
properties of the
double layer systems in the long wavelength limit, and they agree extremely
well with
the results of the microscopic calculations based on Halperin's wave
function. At the
level of the static correlation functions, the degree of agreement is
similar to the
single layer case\cite{zhk,zhang}. However, there is some discrepancy in
the collective
mode spectrum\cite{fsumrule}
between the CSLG theory and the microscopic theory obtained
using the projected single mode approximation. Within the CSLG theory, the
collective mode
spectrum can be obtained by studying the Gaussian fluctuations
of the bose order
parameter, and one obtains the following spectrum for the in-phase and the
out-of-phase collective modes
\begin{equation}
\omega_+ = \hbar \omega_c \ \ \ ; \ \ \ \omega_- =
\hbar \omega_c \frac{m-n}{m+n} ,
\label{cslg-modes}
\end{equation}
in the long wavelength limit.
One sees that the in phase mode agrees exactly with
the prediction of the Kohn's theorem, since the in-phase magnetic
translation is
a good symmetry. The out-of-phase mode agrees exactly with the SMA
spectrum obtained using the {\it full density operator $\rho_q$}
\begin{equation}
\frac{<\rho_q [ H, \rho_{-q}]>}{<\rho_q \rho_{-q}>}
= \hbar \omega_c \frac{m-n}{m+n}.
\label{full-sma}
\end{equation}
This result is not surprising, since this SMA formula involves only the static
correlation functions of the full density operator, and these are obtained
correctly within the CSLG theory, as shown above. The drawback of the CSLG
theory
lies in the fact that there is no sensible way of obtaining the projected
density
operators. In the limit of large Landau level spacing, the SMA using the full
density operator is not adequate, and the projected operators must be used.
Explicitly projecting onto transitions from the $N=0$ to the
$N=1$ Landau levels, one obtains\cite{fsumrule}
\begin{equation}
\omega_+ = \hbar\omega_c \ \ \ ; \omega_- = \hbar\omega_c - \int {d^2 q
\over (2\pi)^2} q^2 V_q^{E} h^{E}(q).
\label{proj-sma}
\end{equation}
for the inter-Landau level transition modes.  From
a microscopic point of view, this discrepancy should be resolved. However,
one can also take a phenomenological point of view, and fix the mode
frequencies
obtained within the CSLG theory as parameters fitted to the projected SMA
calculations.

It is clear for the case $m=n$ that the above formulation breaks down.
The true wave function is not simply given by Eq.(\ref{wavefunction})
which lacks all correlations in the $\pi^-$ channel.  Instead one must
build in Jastrow correlation of the form discussed in Section
\ref{sec:colmodes} in order
to obtain the correct linear dispersion of the long wavelength spin
fluctuations.

For the special case of the $(mmm)$ states, the $SU(2)$ symmetry of the
problem has been exploited by Lee and Kane\cite{leekane},
and they formulated a slightly
different version of the Chern-Simons theory that is very useful for
uncovering the spin charge connection. Their original motivation was to
understand the spin unpolarized quantum Hall effect. However, with a
simple change from spin to pseudospin, one can easily apply this
formalism to double-layer systems (in the zero-separation limit where
the SU(2) symmetry is preserved).  
 In the case of the $(mmm)$ states,
one only needs to introduce a single gauge field, and the action is given by
\begin{eqnarray}
{\cal L}_\phi & = & \phi\sp\dagger_{\sigma}
 (i \partial_t + A^0 -  a^0) \phi_{\sigma}
  - \frac{1}{2m^*} \left|\left(\frac{1}{i} \; {\bf\nabla}
  + {\bf A} - m {\bf a} \right) \phi_\sigma \right|^2 + \frac{1}{4\pi} \;
\varepsilon\sp{\mu\nu\rho} \; a^\mu \partial_\nu \; a^\rho
\nonumber \\
  & & \mbox{} - \frac{1}{2} \int d^2y \; \delta\rho_{\sigma}(x) \; V_0 (x-y) \;
  \delta\rho_{\sigma}(y).
\label{lee-kane}
\end{eqnarray}
In this case, the mean field equations are given by
\begin{equation}
{\bf\nabla} \times \; {\bf a}({\bf x}) =
2\pi \rho_{\sigma}({\bf x}) \ \ \ ; \ \ \
{\bf A} =  m {\bf a}
\label{lk-mean-field}
\end{equation}
and one sees easily that these equations are satisfied when the filling
fraction is $\nu=1/m$. We now decompose the bose fields in the form
\begin{equation}
\phi_\sigma = \sqrt{\bar{\rho}+\delta\rho_{\sigma}} \; \phi \; z_\sigma\ \
\ ; \ \ \
\bar\phi \phi = \bar z_\sigma z_\sigma = 1.
\label{lk-decompose}
\end{equation}
Here $\phi$ is a complex number of unit magnitude and $z_\sigma$ represents
the spinor variable and is related to the unit vector order parameter
$\bf m$ defined previously by
${\bf m}= \bar z_\mu \; {\bf\sigma}_{\mu\nu}\; z_\nu$.

Performing the standard duality transformation\cite{leekane}, one obtains
the following effective action in the dual representation
\begin{eqnarray}
{\cal L} & = &  2\pi \; b_\alpha \; (J_\alpha^v + J_\alpha^s) +
  \frac{m^*}{2\bar{\rho}} \; (\partial_0 \; b_\alpha )^2 \nonumber \\
  & & \mbox{} - \frac{1}{4\pi} \frac{\bar{\rho}}{m} \left[
  2\pi(J^v_0 + J^s_0) - \frac{2\pi}{\nu} \; \varepsilon^{\alpha\beta} \;
  \partial_\alpha \; \delta b_\beta \right] \;
  \ln|x-y|\;
  \left[2\pi(J^v_0 + J^s_0) - \frac{2\pi}{\nu} \;
  \varepsilon^{\alpha'\beta'} \; \partial_{\alpha'}
  \; \delta b_{\beta'} \right]  \nonumber \\
  & & \mbox{} - \frac{1}{2} \;
  (\varepsilon^{\alpha\beta} \; \partial_\alpha \; \delta b_\beta ) \;
  V(x-y)(\varepsilon^{\alpha'\beta'} \; \partial_{\alpha'} \;
  \delta b_{\beta'}),
\label{dual}
\end{eqnarray}
where $\delta b_\alpha \equiv b_\alpha + \frac{\nu}{2\pi}A_\alpha$ and
the gauge field $b_\mu$ is defined by
\begin{equation}
J_{\mu} = \epsilon_{\mu\nu\rho}\partial_\nu b_\rho
\end{equation}
with $J_\mu$ being the three current of the fluid,
\begin{equation}
J^v_{\mu} = \frac{1}{2\pi i} \epsilon_{\mu\nu\rho} \partial_\nu (\bar\phi
\partial_\rho \phi)
\end{equation}
is the vortex three current and
\begin{equation}
J^s_{\mu} = \frac{1}{2\pi i} \epsilon_{\mu\nu\rho} \partial_\nu (\bar
z_\sigma\partial_\rho z_\sigma)
\end{equation}
is the Skyrmion three current.
Note that the word vortex in this context refers to vortices in the
bosonic Chern-Simons field, and should not be confused with vortices
in the spin field discussed in the other sections in terms of the
pseudospin XY ferromagnet analogy.

{}    From this dual action, several important results follow. First of all,
one sees that there is a long ranged logarithmic interaction of the
topological density with itself $\rho_{\rm top}(r)= J_0^v + J_0^s -
\frac{1}{\nu}\varepsilon^{\alpha\beta} \;
\partial_\alpha \; \delta b_\beta $.
In the low energy sector, the only
excitations which can be created are those which have no net topological
charge,
i.e. $\int d^2 r \rho_{\rm top}(r) =0$.
Therefore, these elementary excitations
can be classified into three categories. A vortex excitation has
\begin{equation}
J_0^v = \frac{1}{\nu}
\varepsilon^{\alpha\beta} \; \partial_\alpha \; \delta b_\beta.
\label{vortex}
\end{equation}
Therefore, it carries charge $\pm \nu e$ depending on the sign of the
vorticity.
A Skyrmion excitation has
\begin{equation}
J_0^s =  \frac{1}{\nu}
\varepsilon^{\alpha\beta} \; \partial_\alpha \; \delta b_\beta
\label{skyrmion}
\end{equation}
and it also has charge $\pm \nu e$,
depending on the sign of the Skyrmion number.
This is the equation in the Chern-Simons theory which relates the spin
and charge, which was first noted by Sondhi et al\cite{sondhi}. We see that
this relation is exactly the same as the one obtained in
Eq.(\ref{eq:3.240})
from a microscopic calculation.
One can also form a bound state between these objects so that the net
charge is zero.
\begin{equation}
J_0^s = -J_0^v
\label{bound}
\end{equation}
In this case, the vorticity cancels the Skyrmion number exactly, these
objects are neutral.

The statistics of various spin textures can also be worked out explicitly
from the dual action, following an approach used by Lee and
Zhang\cite{leezhangdual} in the case of the CSLG theory for the
single-layer QHE.  From Eqs.(\ref{dual}), one sees that the first term
$2\pi b_\alpha(J^v_\alpha +J^s_\alpha)$
couples the Skyrmion density $J^s_0$ to the
dual gauge filed $b_\alpha$.  The coupling to the averaged flux $\langle
b_\alpha\rangle$ gives the dynamics of the spin degrees of freedom discussed
below.  There is also a coupling to the fluctuating part, $\delta b_\beta$,
which is given by Eq.(\ref{skyrmion}).  Therefore, this gives rise to a
statistical interaction between the Skyrmion density and the Skyrmion
current of the usual form
\begin{equation}
\nu \int\, d^2x\, d^2y J^s_0 \epsilon^{\alpha\beta} \frac{x^\alpha -
y^\alpha}{|{\bf x-y}|^2} J^s_\beta({\bf y}).
\end{equation}

Another important consequence of the dual action in Eq.(\ref{dual})
 is the form
of the effective spin action. In terms of the $CP^1$ fields, it is given by
\begin{equation}
{\cal L} = i \bar\rho\; (\bar z_\sigma \partial_t z_\sigma) - \frac{K}{2}
(|\bf \partial z_\sigma|^2 + (\bar z_\sigma \bf\partial z_\sigma)^2)
-i J_\mu (\bar z_\sigma \partial_\mu z_\sigma).
\label{spin-action}
\end{equation}
Without the last term which couples the spin and charge degrees of freedom,
this action can be transformed into the angular variables
${\bf m}= \bar z\; {\bf\sigma}\; z$, and its form agrees exactly
with the effective spin action in Eq.(\ref{monopole_action})
derived earlier from the microscopic calculation.
The microscopic calculation was carried out with the assumption that
the charge degrees of freedom is massive. Under this assumption, integrating
out the charge degrees of freedom in Eq.(\ref{spin-action}) will only
produce a long ranged Coulomb interaction between the topological density.
However, strictly speaking, this assumption is true only at zero temperature.
Eq.(\ref{spin-action})
is more generally valid even at finite temperature where the
charge degrees of freedom is gapless. The effect of the gapless charge degrees
of freedom on the spin dynamics and the Kosterlitz-Thouless transition is still
to be explored.
At
this level, the coefficient $K=\bar\rho/m^*$ derived from the
Chern-Simons theory depends on the mass of the electron,
rather than the Coulomb interaction as it should.
This is a general feature encountered
in all Chern-Simons theories. One can view this coefficient as a
parameter and argue that higher-order corrections will bring it into
agreement with microscopic theories. The coefficient of the time dependent
term is independent of the mass, and agrees exactly with the result of
microscopic calculations. At zero temperature, the charge degrees of freedom
have a gap, integrating them out would only give rise to a higher
derivative coupling between the spin variables. Therefore, at least for zero
temperature, the effective spin action given here is sufficient.

Finally we note that, as the layer separation exceeds the critical value
$d^*$, quantum disordering will cause merons to proliferate.  The
path integral
configurations for the meron world lines are very similar to those of the
vortices which disorder the 3D XY model.  However the universality class
of the transition is different because of the coupling to the Chern-Simons
field.\cite{dstar}

\section{Summary}
\label{sec:summary}

We have presented here a theory of the spontaneous development of
interlayer phase coherence in double-layer quantum Hall systems at
various filling factors.  Using a pseudospin language we have shown that the
system is equivalent to an easy plane itinerant ferromagnet with
an unusual spin-charge connection.  There is a zero-temperature
phase transition to a quantum disordered phase if the layer separation
exceeds a critical value $d>d^*$.  For $0<d<d^*$, the system is predicted
to exhibit a finite temperature Kosterlitz-Thouless transition, in the
absence of interlayer tunneling.  Our theory is expected to apply to any
filling factor at which there is an incompressible state which is not
a pseudospin singlet.  Here however we have concentrated primarily on the
case of filling factor $\nu = 1$.

In a companion paper\cite{II} we will
discuss the new phase transitions that occur in the presence of tunneling
and parallel magnetic field.  We will also make contact with the recent
experiments of Murphy et al.\cite{murphyPRL} which appear to have observed
one of these phase transitions.

\section{Acknowledgments}
\label{sec:ack}

It is a pleasure to acknowledge useful conversations with Daniel Arovas,
Greg Boebinger, Nick Bonesteel, Luis Brey, Ren\'e C\^ot\'e, Jim Eisenstein,
Herb Fertig, Matthew Fisher,
Jason Ho, Jun Hu, David Huse, Sheena Murphy, Phil Platzman, Scott Renn,
Nick Read, Ed Rezayi, Mansour Shayegan, Shivaji Sondhi, and Mats Wallin.
The work at Indiana University was supported by
NSF DMR91-13911.  The work at the University of Kentucky was supported
by NSF DMR92-02255.  We are pleased to acknowledge the Aspen Center for
Physics where part of this work was performed.

\newpage
\section{Appendix: Gauge Invariance and Uniform Currents in Landau Levels}
\label{sec:append}

The (static) effective action we have derived for the case of easy plane
anisotropy has as its leading gradient term
\begin{equation}
S = \frac{1}{2} \rho_E |{\bf \nabla}\varphi|^2.
\end{equation}
The phase $\varphi$ describes the local spinor orientation
\begin{equation}
\left(\begin{array}{c} e^{-i\varphi/2} \\ e^{+i\varphi/2}
\\ \end{array} \right),
\end{equation}
and the `charge'  conjugate to $\varphi({\bf r})$ is $S^z({\bf r})$,
which gives the local (physical) charge difference between the layers.
In order to study the gauge symmetry in this problem, it is convenient to
introduce charge and pseudospin gauge fields
\begin{equation}
{\bf A_\pm} = {\bf A_\uparrow} \pm {\bf A_\downarrow},
\end{equation}
where ${\bf A_\uparrow}$ and ${\bf A_\downarrow}$
are the electromagnetic vector
potentials in each of the layers.  The order parameter $\varphi$ is gauge
neutral with respect to ${\bf A_+}$ since it corresponds to the condensation of
a (physical) charge-neutral operator
\begin{equation}
e^{i\varphi({\bf r})} \propto \left\langle \psi^\dagger_\uparrow({\bf r})
\psi_\downarrow({\bf r})\right\rangle.
\end{equation}
However a gauge change
\begin{equation}
{\bf A_-} \longrightarrow {\bf A_-} + 2 \frac{e}{\hbar c} {\bf \nabla}\chi_-
\end{equation}
modifies the wave function
\begin{equation}
\left(\begin{array}{c} \psi_\uparrow \\ \psi_\downarrow
\\ \end{array} \right) \longrightarrow
\exp\left(i 2 \chi_- S^z\right)\left(\begin{array}{c} \psi_\uparrow
\\ \psi_\downarrow
\\ \end{array} \right),
\end{equation}
which means that the action has the usual minimal coupling form
\begin{equation}
S = \frac{1}{2} \rho_E |{\bf \nabla}\varphi + \frac{e}{\hbar c}{\bf A_-}|^2,
\end{equation}
and the pseudospin current is given (for ${\bf A_-} = {\bf 0}$) by
\begin{equation}
J_{zz} = { 2 \rho_E \over \hbar } {\bf \nabla \phi}
\label{eq:spincurrentagain}
\end{equation}
which is identical to the result derived in Eq.(\ref{eq:spincurrent})
using the equation of motion for the projected density.  As mentioned
previously in Eq.(\ref{eq:fsumrule}),
the superfluid mode has oscillator strength proportional to
$q^2$ as expected for an ordinary superfluid.  However the coefficient is
proportional to $\rho_E$ and hence non-universal.

The minimal coupling argument given above is quite correct, however there
are remarkably confusing subtleties lurking just beneath the surface.  For
${\bf \nabla \phi}$ equal to a constant, Eq.(\ref{eq:spincurrentagain})
implies the existence of a uniform zero wavevector current in the LLL.
This is paradoxical as we now discuss.

The question of the form of the current operator in the LLL is a subtle one
which has been considered by several
authors.\cite{GMP,sondhikivcurrent,stonecurrent,raja,sondhirajacurrent}
The difficulty lies in the fact that the true unprojected current operator
has no matrix elements within the LLL.  It is purely off-diagonal, taking
for example, the $N=0$ Landau level into the $N=1$ Landau level.  Hence it
would appear to be impossible to have low energy currents at zero wave
vector, despite our previous derivation using the fact that the projected
densities do not commute.  The resolution of this paradox can be seen in
the following simple example.

Consider the SU(2) invariant $\nu=1$ case and let the ground state be fully
polarized up.  Restricting $H$ to the LLL, this state is an exact
eigenstate.  The exact single-magnon excited states
\begin{equation}
\Psi_{\bf k} = \overline{S^-_{\bf k}}\Psi,
\label{smg4again}
\end{equation}
are labeled by a conserved momentum\cite{kallin} due to the fact that
they are gauge neutral (with respect to ${\bf A_+}$.  They correspond to
spin flip magneto-excitons and have a velocity
\begin{equation}
{\bf v} = \frac{1}{\hbar} {\bf\nabla_k}\epsilon_{\bf k} \times {\hat z}
= \frac{1}{\hbar} {\bf\nabla_k}V(\ell^2\bf k) \times {\hat z},
\end{equation}
(where $V({\bf r_1 - r_2})$ is the particle interaction)
and hence would seem to have a finite pseudospin current at $q=\omega=0$.
If we use the true unprojected current operator, then we must take care to
include LL mixing in $\Psi_{\bf k}$.  First order perturbation theory
shows that such mixing is small and proportional to
$\ell|{\bf\nabla} V|/\hbar\omega_c$.  However the current
\begin{equation}
J^\mu_{zz} = \frac{1}{m} \left({\bf \Pi^\mu_\uparrow} -
{\bf \Pi^\mu_\downarrow}\right)
\end{equation}
has matrix elements connecting adjacent Landau levels proportional to
$\ell\omega_c$, so that Landau level mixing, though small, gives a crucial
contribution to the current, independent of the smallness of the mass $m$.
Carrying out the perturbation theory in detail, yields results in
complete agreement (to first order in $\ell|{\bf\nabla} V|/\hbar\omega_c$)
with the minimal coupling considerations and the
equations of motion methods discussed above.

\addcontentsline{toc}{part}{Figure Captions}

\begin{figure}
\caption{Schematic conduction band edge profile for a
double-layer two-dimensional electron gas system.}
\label{fig1}
\end{figure}

\begin{figure}
\caption{Illustration of magnetic translations and phase factors.
When an electron
travels around a paralellogram (generated by $\tau_q\tau_k\tau_{-q}\tau_{-k}$
) it picks up a phase $\phi=2\pi{\Phi\over \Phi_0}=q^k$ where $\Phi$ is the
flux enclosed in the paralellogram and $\Phi_0$ is the magnetic flux quantum.}
\label{fig:magtrans}
\end{figure}

\begin{figure}
\caption{ Anisotropy mass (in unit of $e^2/\epsilon\ell^3$)
and easy-plane spin stiffness (in unit of $e^2/\epsilon\ell$)
as a function of layer separation for $\nu =1$.  These
results do not include quantum fluctuations which
are important at finite layer separation.}
\label{fig:massstiffnu1}
\end{figure}

\begin{figure}
\caption{ Anisotropy mass and easy-plane spin stiffness
as a function of layer separation for $\nu =1/3$.  These
results do not include the dependence of quantum fluctuations
on the layer separation. }
\label{fig:massstiffnu3}
\end{figure}

\begin{figure}
\caption{ Illustration of a skyrmion on a sphere.}
\label{fig:fig41}
\end{figure}

\begin{figure}
\caption{ Illustration of localized quasiparticle and quasihole
excitations at $\nu = 1$.}
\label{fig:fig42}
\end{figure}

\begin{figure}
\caption{ The number of spins flipped in the ground state versus the
system size (Landau
level degeneracy) when a single electron is added to the $\nu=1$ incompressible
state on a torus.
These results demonstrate that the ground state contains a single
skyrmion spin-texture whose size is determined by the competition between
minimizing the frustration required by the boundary condition and the Coulomb
energy.}
\label{fig:sqspin}
\end{figure}

\begin{figure}
\caption{Microscopic skyrmion energy vs. the scale size $\lambda$.
The trial wavefunction interpolates continuously between a
spin-polarized quasihole at $\lambda=0$ and a smooth
skyrmion spin-texture for $\lambda \to \infty$.}
\label{fig:microskr}
\end{figure}

\begin{figure}
\caption{ Illustration of merons (vortices).  The spin configuration
in the core region tips smoothly out of the XY plane making this object
essentially one-half of a skyrmion.}
\label{fig:meron}
\end{figure}

\begin{figure}
\caption{Illustration of a meron pair with opposite vorticity and
like charge.  We propose that under appropriate circumstances these
objects can form the lowest energy charged excitations in the
system.}
\label{fig:mpair}
\end{figure}

\begin{figure}
\caption{Chemical potential discontinuity $\Delta\mu$
as a function of layer separation $d$ for $\nu=1$.
The results are for system sizes of eight and ten electrons.
The `$\ast$' mark shows the value of $\Delta\mu$ at $d=0$ and $N=\infty$
according to the skyrmion theory.}
\label{fig:fz1}
\end{figure}

\begin{figure}
\caption{Ground state energy versus tunneling amplitude
at small tunneling amplitudes at $d/ \ell =0.2$ for
$N=4$, $N=6$ and $N=10$.}
\label{fig:vsN}
\end{figure}

\begin{figure}
\caption{Ground state energy versus tunneling amplitude
at small tunneling amplitude for $N=10$ at $d/ \ell =0$,
$d / \ell = 0.2 $ and $d/ \ell =0.6$.}
\label{fig:vsd}
\end{figure}

\begin{figure}
\caption{Finite size estimate for the ground-state order
parameter as a function of layer separation for $\nu=1$.
The results are for system sizes of six, eight, and ten electrons.}
\label{fig:mag}
\end{figure}

\begin{figure}
\caption{Finite size estimate for the ground-state
magnetic susceptibility as a function of layer separation for
$\nu = 1$.  The results are for system sizes of six, eight, and
ten electrons.}
\label{fig:chi}
\end{figure}

\begin{figure}
\caption{ Dependence of the ground state energy level on
$d/ \ell$ for one and two electrons transferred between
wells.}
\label{fig:beta}
\end{figure}

\begin{figure}
\caption{ Wavevector dependence of low energy excited states
for $\nu=1$ $ \;$,  $d/ \ell =0.5$ and $N=10$.  These results can
be used to estimate the quantum renormalized spin-stiffness.}
\label{fig:colmodea}
\end{figure}

\begin{figure}
\caption{ Wavevector dependence of low energy excited states
for $\nu=1$ $\;$, $d/ \ell =1.0$ and $N=10$.  These results can
be used to estimate the quantum renormalized spin-stiffness.}
\label{fig:colmodeb}
\end{figure}

\begin{figure}
\caption{ Estimate of Kosterlitz-Thouless transition temperature
for $ B = 4.364$ Tesla so that $ e^2/ \ell k_B \sim 106$ Kelvin.
These estimates include quantum renormalizations of the
spin-stiffness and corrections to the XY model due
to finite anisotropy strength.}
\label{fig:tckt}
\end{figure}


\begin{references}
\bibitem{faith} At $\nu=1$ there is only one $S=N/2$ multiplet in the
Hilbert space.  The orbital wavefunction for this multiplet is the
single-Slater determinant wavefunction of a full Landau
level of spinless electrons.  The $S=N/2$ state is expected on physical
grounds to be the ground state for the case of Coulomb interactions
between electrons,
although we are not aware of any mathematically rigorous proof.
Numerical finite-size exact diagonalization calculations
provide unequivocal support for this expectation.
\bibitem{singletcomment} Singlet states occur at some filling
factors because the constraints imposed on correlations by
the requirement of staying in the lowest Landau level can
be more deleterious for spin-polarized states than for singlet
states.
\bibitem{footnote3.1} An alternate way to define the spin texture state
is to project the entire rotation operator to get $\overline{e^{-iO}}$.
It turns out these two
definitions of the spin texture give identical results for the
quantities we calculate.  Our present
approach has the advantage that $e^{-i\overline{O}}$ is unitary.
\bibitem{noscatt} These terms are certainly present but should
be much smaller than the interaction terms we retain
as long as the envelope functions are small in the barriers separating
the two quantum wells.  Such terms can {\it not\/} be neglected in
single wide quantum wells such as those studied by Suen
{\it et al.\/}\onlinecite{expamd}.
\bibitem{thickness} The finite thickness of a layer can be accounted
for by including form factors corresponding to the isolated subband
envelope functions.  See for example Ref. \onlinecite{ahmz1}.
\bibitem{tcrmk} Although we have not been able to construct a
rigorous proof we believe that the connection between
topological charge and electron charge, which we have
derived for the case of $SU(2)$ invariant interactions,
is not altered by the quantum fluctuations which reduce the
magnitude of the order parameter for $d/ \ell \ne 0$.
\bibitem{stability} Our expressions of the spin stiffness provide a criterion
for the stability of the ferromagnetic states since the
stiffness must be positive.
For example, in the SU(2) invariant case in a $V_0$, $V_1$ model
we would have to have $V_0 > 3 V_1$ in order to get
a stable ferromagnetic state.
\bibitem{jasonhononlinear}  Jason Ho has recently considered oscillatory
effects due to non-linear response to finite electric fields, oppositely
directed in each layer.  See Ref.[\onlinecite{jasonho}].
\bibitem{sfcaveat} The present system has charged vortex-antivortex
excitations as well as neutral vortex-antivortex excitations.
We have not yet fully analyzed the possible importance of these
excitations for the finite-temperature Kosterlitz-Thouless transition
when disorder is present in the system.
\bibitem{more} Further numerical calculations performed
with a finite tunneling amplitude will be detailed in
Ref.[\onlinecite{II}].
\bibitem{attempt} We have attempted to do a finite-size
scaling analysis of our results with limited success.
\end{references}
\end{document}